\documentclass[twocolumn]{aastex63}
\usepackage{graphicx}
\usepackage{kotex}
\usepackage{color}
\usepackage{amsmath}
\usepackage{preamble}

\newcommand{\pok}{{\rm \, cm^{-3}\,K}}

\newcommand{\nicm}{n_{\rm ICM}}
\newcommand{\dicm}{\rho_{\rm ICM}}
\newcommand{\vicm}{v_{\rm ICM}}
\newcommand{\cicm}{c_{s, {\rm ICM}}}
\newcommand{\Picm}{P_{\rm ICM}}

\newcommand{\Wext}{\mathcal{W}_{\rm ext}}
\newcommand{\Wgg}{\mathcal{W}_{\rm GG}}
\newcommand{\kB}{k_{\rm B}}
\newcommand{\Ha}{H$\alpha$}

\newcommand{\zmin}{z_{\rm min}}
\newcommand{\zmax}{z_{\rm max}}
\newcommand{\ssn}[1][]{f_{\rm SN}^{\rm #1}}
\newcommand{\sicm}[1][]{f_{\rm ICM}^{\rm #1}}

\newcommand{\noicm}{{\tt noICM}}
\newcommand{\icmpww}{{\tt ICM-P1}}
\newcommand{\icmpw}{{\tt ICM-P3}}
\newcommand{\icmps}{{\tt ICM-P7}}
\newcommand{\icmpwh}{{\tt ICM-P3h}}
\newcommand{\icmpsh}{{\tt ICM-P7h}}
\newcommand{\icmpss}{{\tt ICM-P14}}

\shorttitle{Ram Pressure Stripping of the multiphase ISM}
\shortauthors{Choi, Kim, \& Chung}

\begin{document}


\title{Ram pressure stripping of the multiphase ISM: a detailed view from TIGRESS simulations}


\author[0000-0001-5033-7208]{Woorak Choi}
\affil{Department of Astronomy, Yonsei University, 50 Yonsei-ro, Seodaemun-gu, Seoul 03722, Korea}
\email{woorak.c@gmail.com}

\author[0000-0003-2896-3725]{Chang-Goo Kim}
\affiliation{Department of Astrophysical Sciences, Princeton University, Princeton, NJ 08544, USA}
\email{cgkim@astro.princeton.edu}

\author[0000-0003-1440-8552]{Aeree Chung}
\affiliation{Department of Astronomy, Yonsei University, 50 Yonsei-ro, Seodaemun-gu, Seoul 03722, Korea}
\email{achung@yonsei.ac.kr}


\begin{abstract}
Ram pressure stripping (RPS) is a process that removes the interstellar medium (ISM) quickly, playing a vital role in galaxy evolution. Previous RPS studies have treated the ISM as single-phase or lack the resolution and physical processes to properly capture the full multiphase ISM. To improve this simplification, we introduce an inflowing, hot intracluster medium (ICM) into a self-consistently modeled ISM in a local patch of star-forming galactic disks using the TIGRESS framework. Our simulations reveal that the workings of RPS are not only direct acceleration of the ISM by ICM ram pressure but also mixing-driven momentum transfer involving significant phase transition and radiative cooling. The hot ICM passes through the low-density channels of the porous, multiphase ISM, shreds the cool ISM, and creates mixing layers. The ICM momentum is transferred through the mixing layers while populating the intermediate temperature gas and radiating thermal energy away. The mixed gas extends beyond galactic disks and forms stripped tails that cool back unless the ICM fluxes are large enough to prevent cooling until they escape the simulation domain. The mixing-driven momentum transfer predicts that the more ICM mixes in, the faster the ISM moves, resulting in the anti-correlation of outflow velocity and gas metallicity of the stripped ISM. The compression of the ISM disks due to the ICM ram pressure enhances star formation rates up to 50\% compared to the model without ICM. With the ICM ram pressure higher than the disk anchoring pressure, star formation is quenched within $\sim$100 Myr.
\end{abstract}


\section{Introduction} \label{sec:introduction}

The current concordance cosmology predicts that smaller-scale structures form first and large-scale structures form through hierarchical merging \citep[e.g.,][]{1985ApJ...292..371D}. In this scenario, the continuous merging of dark matter halos (and galaxies within them) is the chief channel of galaxy formation and evolution, which involves a variety of interactions of galaxies with each other and with their environments \citep{voit2005_review}. Among many types of interactions, the interstellar medium (ISM) in galaxies moving through the intracluster medium (ICM) can be removed by the ICM's ram pressure \citep{1972ApJ...176....1G}. This process, so-called ram pressure stripping (RPS), has been extensively studied due in part to its unique observational signatures identified by disturbed gaseous medium with undisturbed stellar components of galaxies \citep[see][for reviews]{boselli2014_review,boselli2021_review}. Also, RPS is known to effectively affect the ISM content of galaxies on a relatively short time scale \citep{abadi1999,boselli2009}, dramatically changing galaxy evolution in high-density environments.

The observational signatures of RPS are traced in multi-wavelengths. The atomic hydrogen (\ion{H}{1}) 21~cm line has been a pioneering tool to identify RPS galaxies since \ion{H}{1} gas is generally diffuse and extended well beyond the stellar disk, and hence vulnerable to the interaction with the surroundings. The early single-dish observations such as \citet{1984AJ.....89..758H} found the cluster population to be overall deficient in \ion{H}{1} compared to the field counterpart. More direct signatures of RPS such as gas truncation into the stellar disk and/or gas tails have been reported by many \ion{H}{1} imaging studies \citep[e.g.,][]{1990AJ....100..604C,2004AJ....127.3361K,2009AJ....138.1741C,2010MNRAS.403.1175S,2020A&A...640A..22R}. RPS galaxies also show common signs including locally enhanced synchrotron radiation or ionized gas tails which can be observed through radio continuum, optical lines, or X-ray emission \citep[e.g.,][]{gavazzi2001,2010A&A...512A..36V,2017ApJ...840L...7S,sun2010,poggianti2017_gasp_muse}.

On the other hand, the case of molecular gas content, which is generally present in the inner part of galaxies with a higher density compared to the other ISM phases, does not show clear evidence of RPS \citep[e.g.,][]{1989ApJ...344..171K}. Depending on the sample selection and the observational strategy, only a handful number of studies find molecular gas deficiencies \citep[e.g.,][]{2009ApJ...697.1811F,corbelli2012,boselli2014_molecule_def,2017ApJ...843...50C}, and the impact of ram pressure on the molecular gas disk has long been under debate. However, more recent high-resolution radio observations begin to show that the morphological characteristics of RPS seen in \ion{H}{1} disks such as asymmetry and compression are shared by CO disks \citep[e.g.,][]{2017MNRAS.466.1382L,2018ApJ...866L..10L,2021ApJS..257...21B} and have revealed the enhancement of molecular gas \citep[e.g.,][]{moretti2018_molecule,moretti2020,moretti2020b,cramer2021}. These data imply that the molecular ISM is essentially affected by ram pressure in similar ways as more diffuse components although whether it is stripped along with atomic gas or not is still unclear. All these observations indicate that RPS is the process in which the multiphase ISM from cold molecular gas to hot ionized gas is involved. 

This naturally raises a question on the effect of RPS in star formation, whether RPS is only effective in stripping the diffuse gas limiting the supply of gas for future star formation \citep[e.g.,][]{2004ApJ...613..851K,2008AJ....136.1623C} or even enhance star formation by compressing the gas further \citep[e.g.,][]{merluzzi2013_sfr,2014ApJ...780..119K,2018ApJ...866L..25V}. There have been extensive observational studies showing that both scenarios are possible.
Therefore more complete understandings of the impact of ram pressure and its consequences on star formation and galaxy evolution require the studies of the multiphase ISM responding to the ICM ram pressure.

A large number of theoretical studies of RPS have been conducted mainly using numerical simulations in two contexts. 
One is self-consistent cosmological simulations within which galaxies experience RPS as they are moving in clusters \citep[e.g.,][]{2016A&A...591A..51S,2017MNRAS.468.4107R,2018ApJ...865..156J,yun2019_TNG}, and the other is more controlled simulations of a single galaxy interacting with an inflowing ICM \citep[so called wind-tunnel simulations, e.g.,][]{2006A&A...453..883V,2008A&A...481..337K,2009A&A...500..693J,2009ApJ...694..789T,2010ApJ...709.1203T,2012A&A...544A..54S,2014ApJ...795..148T,2014ApJ...784...75R,2018MNRAS.476.3781R,2020ApJ...905...31L,2021ApJ...911...68T}.  
Such simulations reproduce long tails seen in observations and show overall agreement with the prediction for the effectiveness of RPS when the ICM ram pressure is stronger than the ISM anchoring pressure \citep[e.g.,][]{2007MNRAS.380.1399R,2009ApJ...694..789T}. 
In both cases, the outer dimensions covered had to be larger than a few tens of kpc, banning the use of pc-scale, high-resolution required for explicit modeling of the ISM physics. 
Instead, subgrid models of multiphase ISM, star formation, and feedback \citep[e.g.,][]{2003MNRAS.339..289S} are often adopted.  

There exists a few RPS galaxy simulations that include gas cooling down to $\sim 100$~K, but most of them has not particularly focused on modeling of the full multiphase (cold, warm, and hot) ISM properly \citep[e.g.,][]{2010ApJ...709.1203T,2012MNRAS.422.1609T,2021ApJ...911...68T,2014MNRAS.438..444B}. The radiative heating by photoelectric effect of FUV on small grains is ignored, which is the major heating source of the warm and cold ISM \citep{1995ApJ...443..152W,2003ApJ...587..278W}. Simulations with not enough resolution \citep{2012MNRAS.422.1609T,2014MNRAS.438..444B} cannot resolve the Sedov-Taylor stage of SNe that are critical in driving turbulence and creating the hot gas \citep{2015ApJ...802...99K,2017ApJ...834...25K_KOR17,steinwandel2020}. Such simulations tend to overcooling the gas and confine the ISM in very thin, unresolved disks. The mass and volume distributions of the multiphase ISM that the ICM is interacting are severely compromised. To our best knowledge, only \citet{2020ApJ...905...31L} have marginally high resolution and set of physics to treat the full range of the multiphase ISM and star formation and feedback explicitly.

Nevertheless, \citet{2021ApJ...911...68T} conducted the RPS simulation with a single galaxy with a full range of cooling function and claimed that RPS occurs via mixing between the ICM and ISM. 
This interesting result qualitatively agrees with the recent in-depth studies of radiative mixing layers \citep{2020ApJ...894L..24F,2021MNRAS.502.3179T} in context of shock/wind-cloud interaction simulations \citep{2018MNRAS.480L.111G,2020MNRAS.492.1970G,2021arXiv210713012G,2020MNRAS.492.1841L,2021MNRAS.501.1143K,2020MNRAS.499.4261S,2021arXiv210110344A} and starburst-driven galactic winds \citep{2020ApJ...895...43S} that emphasize the mixing-driven momentum transfer as the major acceleration mechanism for cooler, denser gas. 

To enable global galaxy simulations with a large dynamic range, the usual practice is to adaptively refine the resolution elements to achieve constant mass resolution \citep{2020ApJ...905...31L,2021ApJ...911...68T}.
Given typical $\sim2$ decades temperature contrast between cold, warm, and hot phases in pressure equilibrium, the spatial resolution of adjacent thermal phases differs by a factor of 5. 
The interaction between hot and cooler phases and the mixing layers produced by such interaction can be severely altered by large differences in spatial resolution of interacting phases.
Simulations with uniformly high resolution are thus necessary to model different phases and their interactions more robustly \citep[e.g.,][]{2018ApJ...853..173K,2020ApJ...895...43S}.
Since the mixing-driven momentum transfer is key physical process in general multiphase hydrodynamical interactions, the multiphase RPS deserves more careful studies using self-consistent, multiphase ISM models that interacts with the ICM.

To this end, we conduct a new suite of numerical simulations focusing on a smaller section of galactic disks with an inflowing ICM. Our numerical models build on the TIGRESS framework developed to model the star-forming ISM self-consistently \citep{2017ApJ...846..133K}. TIGRESS solves ideal magnetohydrodynamics (MHD) in a shearing-box with Athena \citep{2008ApJS..178..137S} and includes additional ISM physics including optically-thin cooling at full temperature range, self-gravity of gas and newly formed stars, star cluster formation in gravitationally bound objects using sink particles, and massive star feedback in the forms of supernovae (SNe) and far-ultraviolet (FUV) radiative heating. The original closed box model has been used to study the internal regulation of star formation rates (SFRs) and driving of multiphase outflows \citep{2020ApJ...900...61K,2018ApJ...853..173K,2017ApJ...846..133K} among other implications. In this paper, we choose a particular ISM model representing the solar neighborhood condition. We take a snapshot of self-consistently modeled multiphase ISM in a quasi-steady state and conduct controlled numerical experiments with different ICM ram pressures covering relatively weak and strong pressure regimes compared to the ISM anchoring pressure by stellar disks. Our chosen parameters approximately represent the conditions of NGC~4522, a prototypical RPS galaxy in Virgo, at different radii \citep{2004AJ....127.3361K}. 

As the first of a kind study using local models, this paper focuses on fostering an in-depth understanding of the inner workings of \emph{multiphase} RPS. In addition, with the help of self-consistent star formation and feedback models of TIGRESS, we investigate how RPS affects star formation in and out of the galactic disks. In the future, we further study the role of magnetic fields in RPS, especially on the dense, molecular gas. 

The structure of this paper is as follows. In \autoref{sec:method}, we summarize the TIGRESS framework and introduce the ISM and ICM models. In \autoref{sec:overall}, we first overview our simulations using the time evolution of horizontally-averaged and globally-integrated quantities. \autoref{sec:morphology} then delineate a variety of physical properties in two representative models. In \autoref{sec:stripping_transfer}, we analyze the mass, momentum, and energy transfers between thermal phases. We then check the prediction of the mixing-driven momentum transfer in \autoref{sec:stripping_mixing}. \autoref{sec:sfr} presents the impact of RPS on SFRs and the extraplanar star formation. \autoref{sec:discussion} discusses the main observational imprints from RPS by the mixing-driven momentum transfers. Also, we discuss our results in context and caveats. Finally, the main conclusions are summarized in \autoref{sec:conclusion}.

\section{Numerical Methods and Models} \label{sec:method}

In this section, we begin by summarizing the numerical methods employed in the TIGRESS framework to simulate the multiphase ISM with star formation and feedback in \autoref{sec:method_tigress} for completeness. We then explain the evolution of the ISM without ICM inflows in \autoref{sec:method_ism}. The readers who are familiar with the local ISM simulations and only interested in the ICM-ISM interaction can skip the first two subsections. \autoref{sec:method_icm} explains the ICM inflow setup. The tracer fields and gas phases used throughout the paper are defined in \autoref{sec:method_phase}.

\subsection{TIGRESS framework}\label{sec:method_tigress}

We use the TIGRESS framework developed by \citet[][]{2017ApJ...846..133K} to evolve the multiphase, turbulent, magnetized ISM with which the ICM interacts. We refer the reader to \citet{2017ApJ...846..133K} for full details of the methods and tests. TIGRESS solves the ideal MHD equations in a local shearing-box representing a $\sim$kpc patch of differentially rotating galactic disks using the Athena code \citep{2008ApJS..178..137S,2009NewA...14..139S}. Local Cartesian coordinates $x$ and $y$ correspond to the local radial and azimuthal directions of global galactocentric coordinates such that $(x,y)=(R-R_0, R_0[\phi - \Omega_0 t])$, while $z$ is the vertical coordinate. The simulation domain is corotating with galactic rotation speed at the center of the simulation domain, $\Omega_0\equiv\Omega(R_0)$, arising inertial forces including the Coriolis force and the tidal potential in the momentum equation. The flat rotation curve is assumed, $d\ln\Omega/d\ln R=-1$. We adopt shearing-periodic boundary conditions in the horizontal directions \citep{2010ApJS..189..142S} and outflow boundary conditions in the vertical directions. The bottom vertical boundary conditions are modified for the ICM inflows (see \autoref{sec:method_icm}).

We solve Poisson's equation to obtain gravitational potential from gas and newly formed young stars using the FFT method with horizontally shearing-periodic and vertically open boundary conditions \citep{2001ApJ...553..174G,2009ApJ...693.1316K}.
The gravitational potential of old stellar disks and dark matter halos is held fixed and only exerts vertical gravity. We introduce a sink particle when a gas cell is experiencing unresolved self-gravitating collapse as indicated by the Larson-Penston density threshold at $\rho_{\rm LP} \equiv (8.86/\pi) c_s^2/G\Delta x^2$, where $c_s\equiv (P/\rho)^{1/2}$ is the local sound speed, and $\Delta x$ is the side length of a cubic grid cell used in the simulation \citep{2013ApJS..204....8G}. We adopt additional criteria for the sink particle creation including a converging flow check (in all three directions) and a local potential minimum check. Typically, $\rho_{\rm LP}\sim 100\pcc$ for $8\pc$ resolution and $\sim 300\pcc$ for $4\pc$ resolution.
Note that the typical mass of sink particles is in a range between a few $10^3\Msun$ to $10^5\Msun$, representing star clusters rather than individual stars. We treat each particle as a population of stars with a fully-sampled initial mass function of \citet{2001MNRAS.322..231K}.

We use the STARBURST99 stellar population synthesis model to obtain SN rate and FUV luminosity for each star cluster \citep{1999ApJS..123....3L}. In addition to clustered SNe occurring at the position of the sink particle, we produce a massless particle with a probability of 50\% for each SN event to model a runaway star ejected from a binary OB star \citep{2011MNRAS.414.3501E}. The total SN rate still matches that of the original STARBURST99 model.

For each SN event, we first identify the cells with distances from the explosion center smaller than $R_\mathrm{SNR}=3\Delta x$ and calculate the total mass $M_{\rm SNR}$ and volume $V_{\rm SNR}$ of the feedback region (or the SN remnant). If $M_{\rm SNR}/M_{\rm sf} <1$, where $M_{\rm sf} = 1540\Msun (n_{\rm amb}/\pcc)^{-0.33}$ is the shell formation mass at a given ambient medium density $n_{\rm amb}=M_{\rm SNR}/V_{\rm SNR}$ \citep{2015ApJ...802...99K}, we inject $10^{51}\erg$ dividing into thermal and kinetic energy with the Sedov stage energy ratio of $0.72:0.28$. Otherwise, we inject the terminal momentum of SNR $p_{\rm SNR}=2.8\times10^5\Msun\kms (n_{\rm amb}/\pcc)^{-0.17}$ as calibrated in \citet{2015ApJ...802...99K}. The total and metal mass of SN ejecta, $M_{\rm ej}= 10\Msun$ and $Z_{\rm SN}M_{\rm ej}=2\Msun$ with $Z_{\rm SN}=10Z_\odot$, are traced using passive scalars. See \autoref{sec:method_phase} for details.

We use the total FUV luminosity from star clusters in the simulation domain to set the instantaneous photoelectric heating rate by interstellar radiation field \citep{1994ApJ...427..822B,2001ApJS..134..263W}. We apply the mean attenuation factor using the plane-parallel approximation as in \citet{2020ApJ...900...61K}. As a result, the heating rate varies in time self-consistently but is spatially constant. 

Optically-thin cooling is included in the energy equation using a tabulated cooling rate coefficient $\Lambda(T)$ from \citet{2002ApJ...564L..97K} at $T<10^{4.2}\Kel$ and \citet{1993ApJS...88..253S} at $T>10^{4.2}\Kel$ (collisional ionization equilibrium at solar metallicity is adopted) depending only on temperature. Although we follow the metallicity of gas in each cell (see \autoref{sec:method_phase}), we note that we do not use the metallicity information to set the cooling rate. 

More self-consistent treatment of radiation and chemistry and hence cooling and heating rates is being developed for the TIGRESS framework (J.-G. Kim et. al in prep.), which will enable further study of RPS with more realistic ISM. This extension is particularly important to pursue the extraplanar molecular gas in RPS galaxies.

\subsection{ISM disk model}\label{sec:method_ism}

In this work, we make use of the solar neighborhood model of the TIGRESS simulation suite, setting the parameters for gravitational potential of stars and dark matter (see below; \autoref{eq:phi_ext}). We adopt the angular velocity of galactic rotational $\Omega_0=28\kms\kpc^{-1}$, giving rise to the orbit time $\torb = 2\pi/\Omega_0 = 224\Myr$. We use a vertically elongated rectangular box with the outer dimensions of $(L_x, L_y, L_z)=(1024, 1024, 7168) \pc$. A uniform, cubic grid cell is used with the side length of $\Delta x=8\pc$ at which we achieve convergence of overall properties of the ISM, SFR, and outflows \citep[see][]{2017ApJ...846..133K,2018ApJ...853..173K,2020ApJ...900...61K}. This model is referred to as the \noicm{} model throughout the paper (identical to the solar neighborhood model, R8, presented in other works). Additional details of the solar neighborhood model can be found in \citet{2017ApJ...846..133K} for initial conditions, overall evolution, numerical convergence, and technical details, \citet{2018ApJ...853..173K,2020ApJ...894...12V} for galactic fountains and winds, and \citet{2020ApJ...898...52M} for the properties of gravitationally bound clouds and their connection with SFRs.

The simulation starts from an idealized initial condition with horizontally uniform, vertically-stratified gas profiles with the initial gas surface density of $\Sigma_{\rm gas}=13\Surf$. We introduce initial velocity perturbation and set thermal pressure to ensure that the disk is in rough hydrostatic equilibrium. Soon after the simulation begins, the initially imposed velocity perturbation dissipates, and the gas cools. The overall vertical contraction occurs owing to the reduction of turbulent and thermal pressure, leading to a burst of star formation. SNe and FUV heating from newly formed massive stars respectively offset turbulence dissipation and gas cooling, recovering vertical support against gravity. The disk expands vertically, reducing SFRs and hence feedback. The reduction of feedback causes another disk contraction, and the cycle repeats. Each cycle has a period similar to the vertical oscillation time scales of $~40-50\Myr$ \citep[see][]{2020ApJ...900...61K}. Although the first burst is a consequence of the idealized initial setups, our simulations soon enter a self-consistently regulated state after a few star formation-feedback cycles ($t>100\Myr$ in this model).

\begin{figure}
    \centering
    \includegraphics[width=\columnwidth]{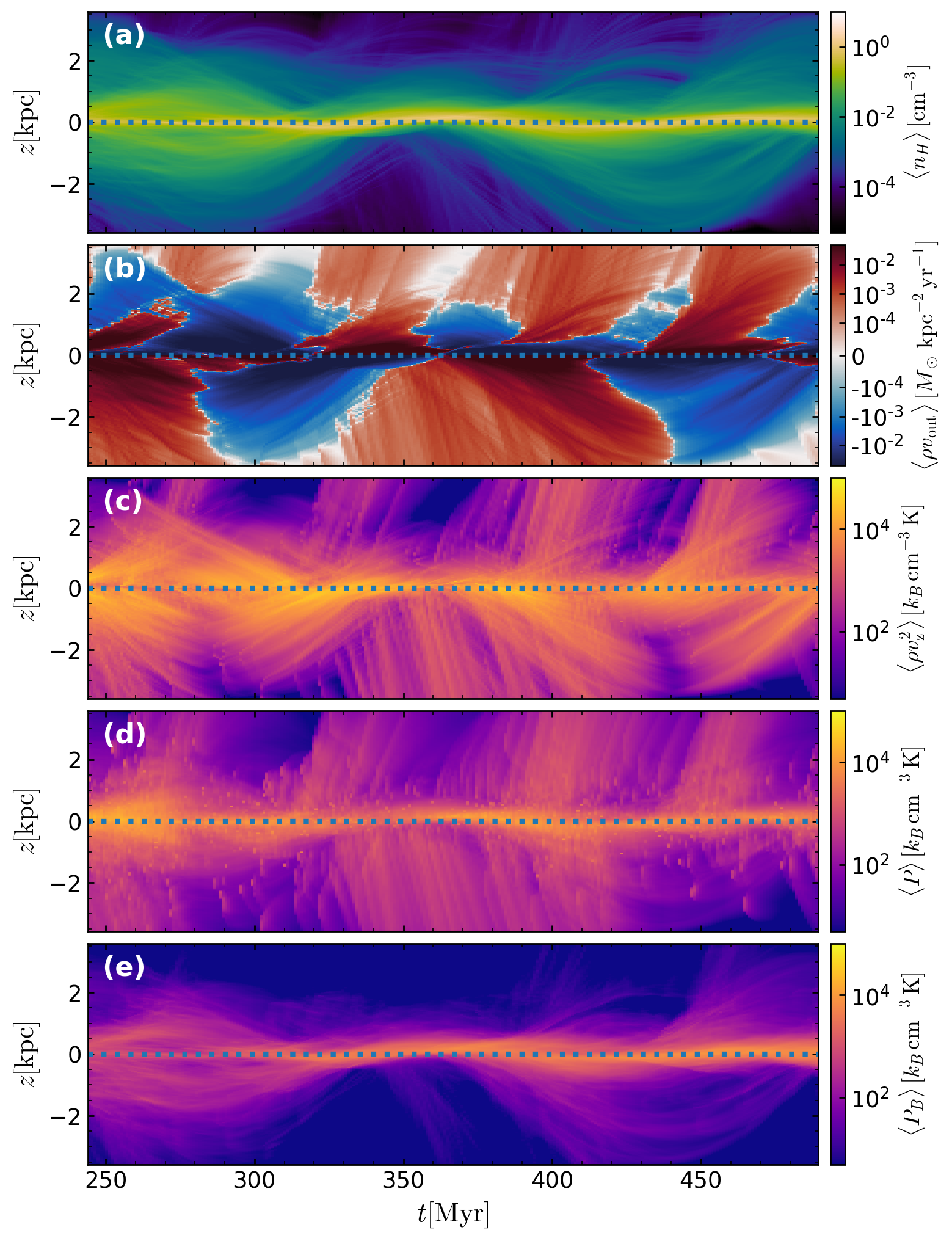}
    \caption{Space-time diagrams of horizontally averaged (a) hydrogen number density $n_H$, (b) outgoing mass flux $\rho v_z {\rm sgn}(z)$, (c) turbulent pressure $\rho v_z^2$, (d) thermal pressure $P$, and (e) magnetic pressure $P_B\equiv B^2/(8\pi)$ for the \noicm{} model. We only show the evolution during a self-regulated state over $t\sim 250-500\Myr$ as a reference that can be directly compared with the models with the ICM. The horizontal dotted line demarks the midplane ($z=0$).}
    \label{fig:noICM}
\end{figure}

To overview the evolution in a quasi-steady state far from the initial burst, \autoref{fig:noICM} shows the horizontally-averaged physical quantities in the space (vertical coordinate $z$) and time ($t$) plane as defined by
\begin{equation}\label{eq:havg}
    \abrackets{q(z;t)}\equiv\frac{\int q(x,y,z;t) dxdy}{L_x L_y}
\end{equation}
for a physical quantity $q$ of interest. We only show a self-regulated state over $t\sim250-500\Myr$.
From top to bottom, we show (a) hydrogen number density $n_H$, (b) outgoing mass flux $\rho v_{\rm out}$, (c) turbulent pressure (or, equivalently, vertical momentum flux) $\rho v_z^2$, (d) thermal pressure $P$, and (e) magnetic pressure $P_B\equiv B^2/(8\pi)$.
Here, the outward vertical velocity is defined by $v_{\rm out} \equiv v_z {\rm sgn}(z)$ such that $v_{\rm out}$ is positive (red) for outflow and negative (blue) for inflow about the midplane. Note that the midplane ($z=0$) in our simulation defines the symmetric plane of the fixed gravitational potential of stars and dark matter. Even without ICM inflows, the gas distribution can be largely asymmetric as stochastic SN explosion cannot be perfectly symmetric. As a result, the total gravitational potential including the gravitational potential of gas and young star clusters can be asymmetric.

Within the simulation duration shown in \autoref{fig:noICM}, we can visually identify four strong outflow launching epochs ($t\sim250$, 320, 380, and 420~Myr; see panel (b)). These epochs are associated with strong star formation events. The outflows in our simulations show clear multiphase nature, consisting of fast, hot winds that escape the simulation domain and slow, warm fountains that fall back to the midplane \citep{2018ApJ...853..173K}. The hot winds ($T\simgt 10^{5-6}\Kel$) can be easily identified by the high thermal and turbulent pressure gas in the extraplanar region $z>1\kpc$ with a steeper slope in the $z$-$t$ plane (panels (c) and (d)). The outgoing mass flux in panel (b) associated with the hot winds is always red (only outflows). However, the warm fountains ($T\simlt 10^{4}\Kel$) are evident with alternating colors in panel (b), implying that the warm outflows are always followed by inflows (see also panel (a)). Only a small fraction of the warm outflow has reached high velocity to escape the simulation domain (see \citealt{2018ApJ...853..173K,2020ApJ...894...12V}). The magnetic pressure in panel (e) is overall subdominant, especially in low-density gas far from the midplane. The magnetic field strength grows over time via galactic dynamo and ranges from a few to ten micro Gauss in the warm and cold medium with comparable turbulent and mean field strengths \citep[see][]{2019ApJ...880..106K}. This is consistent with the observed magnetic field strength of neutral hydrogen in the solar vicinity \citep{2005ApJ...624..773H}. The full complexity of star formation/feedback and multiphase outflow/inflow cycles in the TIGRESS simulation suite is extensively discussed in \citet{2020ApJ...900...61K}.

\subsection{ICM models}\label{sec:method_icm}

\begin{deluxetable*}{lccccc}
\tablecaption{ICM Model Parameters \label{tbl:models}}
\tablehead{
\colhead{Model} &
\colhead{$\nicm$} &
\colhead{$\vicm$} &
\colhead{$\Picm/\kB$} &
\colhead{$\Picm/{\Wgg}$} &
\colhead{$\Delta x$}\\
\colhead{} &
\colhead{($10^{-4}\;\pcc$)} & \colhead{($10^3\;\kms$)} &
\colhead{($10^4\;\pok$)} &
\colhead{} &
\colhead{(pc)}
}
\colnumbers
\startdata
  \icmpww          & 0.5 & 1   & 0.94 & 0.18 & 8 \\
  \icmpw({\tt h})  & 1   & 1.4 & 3.6  & 0.69 & 8(4) \\
  \icmps({\tt h})  & 2   & 1.4 & 7.2  & 1.4  & 8(4)\\
  \icmpss          & 2   & 2   & 14   & 2.7  & 8
\enddata
\tablecomments{
Column (1): model name.
Column (2): hydrogen number density of the ICM.
Column (3): relative velocity of the ICM and the ISM disk.
Column (4): ICM pressure.
Column (5): ratio of the ICM pressure to the ISM weight. $\Wgg=5.27\times10^4\kB\;\pok$ is an approximate ISM weight estimated by \autoref{eq:Wgg}.
Column (6): spatial resolution.
}
\end{deluxetable*}

We take the first snapshot shown in \autoref{fig:noICM} ($t\sim245\Myr$) as initial conditions and restart the simulation with an ICM inflow. Here and hereafter, we exclusively use the term ISM in simulations to denote the gas that was in the simulation domain before injecting the ICM. At this time, gas surface density in the \noicm{} model is reduced to $\Sgas= 9.5 \Surf$ as gas turns into stars and leaves the simulation domain as outflows through the vertical boundaries. We model the ICM inflow as a constant, unmagnetized, vertical inflow through the bottom boundary (i.e., face-on interaction). We set the ICM metallicity to $Z_{\rm ICM} = 0.1Z_\odot$ \citep[e.g,][]{2011MNRAS.414.2101U}, which serves as a tracer of the gas origin together with other passive scalars (see \autoref{sec:method_phase}).


The ICM inflows are characterized by two parameters: hydrogen number density of the ICM $\nicm=\dicm/(1.4271 m_H)$ and inflow velocity $\vicm$. We adopt the ICM sound speed $\cicm=300\kms$. The total
pressure of the ICM at injection is
\begin{align}\label{eq:PICM}
    \Picm \equiv& \dicm(\vicm^2 + \cicm^2) =  \dicm\vicm^2 (1+\mathcal{M}_{\rm ICM}^{-2})\nonumber\\
    \approx& 1.73 \times 10^4 (1+\mathcal{M}_{\rm ICM}^{-2})
    \kB\pok\nonumber\\
    &\left(\frac{\nicm}{10^{-4}\pcc}\right)
    \left(\frac{\vicm}{10^3\kms}\right)^2,
\end{align}
which is dominated by the ram pressure for our chosen $\vicm\ge10^3\kms$ (or Mach number of the ICM $\mathcal{M}_{\rm ICM}>3.3$).\footnote{Note that the adopted ICM sound speed (or $T_{\rm ICM}\sim 4 \times10^6$\Kel) is about a factor of two smaller than that of the ICM in the Virgo cluster ($T_{\rm ICM}\sim 2\times10^7\Kel$; \citealt{shibata2001_virgo_xray}), which is still smaller than the inflow velocity $\vicm$ so that the results are expected to be qualitatively unchanged.} In the simulations, however, as soon as the ICM sweeps up the ISM, a reverse shock thermalizes the inflowing ICM, and it is the hot ICM with the total pressure $\Picm$ dominated by the thermal term that interacts with the ISM. 

While the ISM is pushed away from the galactic disk owing to the interaction with the ICM, the stellar and dark matter components are not immediately disturbed in RPS galaxies. This is particularly true for our simulations because we use a fixed analytic potential for stellar and dark matter gravity. The gas weight under the external gravity\footnote{Note that the term ``external'' here is used not for the gravity from external galaxies but for the gravity from non-gaseous components.} is
\begin{align}\label{eq:Wext}
    \Wext &\equiv \int_0^\infty \! \rho\left|\frac{d\Phi_{\rm ext}}{dz}\right|  \, dz,
\end{align}
where the functional form of the external gravitational potential is
\begin{align}\label{eq:phi_ext}
\Phi_{\rm ext}(z)\equiv &2\pi G \Sigma_* z_*\sbrackets{\rbrackets{1+\frac{z^2}{z_*^2}}^{1/2}-1} \nonumber\\
&+ 2\pi G \rho_{\rm dm} R_0^2\ln\rbrackets{1+\frac{z^2}{R_0^2}}.
\end{align}
We adopt the parameters representing solar neighborhood conditions: galactocentric distance $R_0=8\kpc$, stellar surface density $\Sigma_*=42\Surf$, stellar scale height $z_*=245\pc$, and midplane dark matter density $\rho_{\rm dm}=6.4\times10^{-3}\rhounit$ \citep{2013ApJ...772..108Z,2015ApJ...814...13M}.
When the gas is stripped far away from the stellar disk (i.e., the mean gas position is much larger than the stellar disk scale height, $z\gg z_*$), the stellar gravity (the first term in \autoref{eq:phi_ext}) becomes nearly constant such that $|d\Phi_*/dz| = 2\pi G \Sigma_*$. The ISM weight can then be well approximated by $\Wext\approx\Wgg$, where
\begin{align}\label{eq:Wgg}
    \Wgg \equiv& 2\pi G \Sgas\Sigma_*\\
    =&5.27\times10^4\kB\pok\nonumber\\
    &\rbrackets{\frac{\Sgas}{9.5\Surf}}
    \rbrackets{\frac{\Sigma_*}{42\Surf}}.
\end{align}
This ``restoring'' force per area (often called as ``anchoring'' pressure) originally presented in \citet{1972ApJ...176....1G} has been conveniently compared with the ICM ram pressure to determine the stripping condition \citep[e.g.,][]{2004AJ....127.3361K,2006A&A...453..883V,2007ApJ...659L.115C,koppen2018,jaffe2018}. Note that, for our adopted gravitational potential, the dark matter contribution in the vertical gravity keeps increasing with $z$ and becomes comparable to that of stars at the vertical boundaries $z\sim 3.5\kpc$. Therefore, $\Wgg$ slightly underestimates the maximum $\Wext$ in our simulations by 25\%.

\autoref{tbl:models} lists the ICM models. Column (1) is the model name; we adopt a nomenclature including the strength of the ICM ram pressure presented in Column (4). The higher resolution models ($\Delta x=4\pc$) have their name ending with `{\tt h}' (see Column (6)). For the higher resolution models, we refine the original data cube from the \noicm{} model using a zero gradient prolongation (i.e., volume- and area-averaged quantities in finer grid cells are the same as their parent cell values). Therefore, the initial conditions are identical across all models. Columns (2) and (3) are the number density and inflow velocity of the ICM, which set the ICM pressure (Column (4); \autoref{eq:PICM}). Column (5) shows the ratio of the ICM pressure and the maximum ISM weight under the stellar gravity (\autoref{eq:Wgg}), which is a rough estimate for the relative strength of the ICM pressure to the maximum ISM weight. Finally, we list the spatial resolution in Column (6).

We consider four different ICM conditions (with additional two higher resolution runs), covering $\Picm/\kB \sim 1-14 \times 10^4\pok$. Since the ISM condition is fixed, the relative strength of the ICM-ISM interaction simply increases as $\Picm$ increases; with \icmpww{} and \icmpw{(\tt h)} have $\Picm/\Wgg<1$, and \icmps{(\tt h)} and \icmpss{} have $\Picm/\Wgg>1$. Throughout the paper, the former two models are referred to \emph{the weak ICM models}, and the latter two models are referred to \emph{the strong ICM models}, respectively. Our parameter choice brackets the relative strength of the ISM-ICM interaction seen in NGC 4522, a prototypical galaxy undergoing ram pressure stripping in the Virgo cluster \citep{2004AJ....127.3361K,2006A&A...453..883V}. In addition, the anchoring pressure of our simulation (\autoref{eq:Wgg}) is comparable to that  near the truncation radius of NGC 4522 \citep[][]{2004AJ....127.3361K,2009AJ....138.1741C,2017MNRAS.466.1382L,2018ApJ...866L..10L}. Thus, our \emph{weak/strong} ICM models can represent the evolution of the \emph{inner/outer} part of the truncation radius of NGC 4522.

\subsection{Tracer Fields and Gas Phases}\label{sec:method_phase}

In the TIGRESS framework, the gas is divided into five thermal phases based on its temperature, corresponding typical discriminators of the three-phase ISM (but including thermally unstable phases; \citealt{mckee1977}). Each cell is exclusively assigned as cold, unstable, warm, intermediate (warm-hot ionized medium), and hot phase following the temperature criteria in \autoref{tbl:phase}. We often combine the cold, unstable, and warm phases and call them the cool phase. 

\begin{deluxetable}{lC}
\tablecaption{Definition of Thermal Phases \label{tbl:phase}}
\tablehead{
\colhead{Phase} &
\colhead{Condition} 
}
\startdata
cold & T<184\Kel \\
unstable & 184\Kel<T<5050\Kel  \\
warm & 5050\Kel<T<2\times10^4\Kel  \\
intermediate & 2\times10^4\Kel<T<5\times10^5\Kel  \\
hot & T>5\times10^5\Kel 
\enddata
\tablecomments{The cold, unstable, and warm phases are combined and referred to as the cool phase.}
\end{deluxetable}

The hot gas in the \noicm{} model is created by SN shocks, while the warm and cold phases are maintained via radiative heating due to FUV radiation. With ICM inflows, a significant fraction of the hot ICM is directly added. The hot gas can accelerate the cooler gas directly through its pressure gradient (both ram and thermal pressure), but another significant (likely dominant) acceleration mechanism, as we shall show, is by the momentum transfer through the mixing of the hot gas into the cooler gas (\autoref{sec:stripping}; see also \citealt{2022ApJ...924...82F}). It is therefore critical to separate the origin of gas and trace the fraction of different origin gas in different thermal phases. We utilize passive scalars to track the mass fractions of the initial ISM, SN ejecta, and ICM in each cell. Here, we use the term passive scalar to denote the multiplication of gas density and tracer field (or specific scalar).

Practically, we follow the total metallicity $Z$, SN ejecta mass fraction $\ssn{}$, and ICM mass fraction $\sicm{}$. The metallicity tracer field $Z$ is initialized with $Z_\odot=0.02$ at the beginning of the \noicm{} simulation, while the other two tracer fields are initialized to zero everywhere. In the code, additional continuity equations for $\rho Z$, $\rho \ssn$, and $\rho\sicm$ are solved with the velocity field of gas. For each SN event, we add the total and metal density of SN ejecta, $\rho_{\rm ej}\equiv M_{\rm ej}/V_{\rm SNR}$ and $Z_{\rm SN}\rho_{\rm ej}$, respectively, to passive scalars in the feedback region (of course, the SN ejecta density is added to the gas density as well). As the \noicm{} model has evolved for $\sim250\Myr$ before the restart with the ICM, the ISM disk's metallicity has been enriched by SN ejecta. When we restart simulations with ICM inflows, we adopt metallicity and SN ejecta fraction inherited from the \noicm{} model. The ICM inflow with the ICM tracer field $\sicm=1$ is then added and followed by another passive scalar. Also, the ICM metallicity is set to $Z_{\rm ICM}=0.1Z_\odot$ and mixed into the metal passive scalar $\rho Z$ as the ICM interacts with the existing gas. When sink particles are formed and accrete gas, passive scalars are also locked into particles, which represent the metallicity of star-forming gas.

As mentioned above, we adopt three distinct metallicities for different origin gases:
\begin{itemize}
    \item genuine ISM -- initial gas from the beginning of the \noicm{} model ($Z_{\rm ISM,0}=Z_\odot$),
    \item SN ejecta -- gas added in the feedback region by SNe ($Z_{\rm SN}=10Z_\odot$), and
    \item ICM -- gas added from the bottom boundaries as the ICM inflow ($Z_{\rm ICM}=0.1Z_\odot$).
\end{itemize}
The metallicity is a good proxy for the composition of the gas in simulations, providing potential observational imprints. In each cell, the metallicity is connected to the SN ejecta and ICM mass fractions as
\begin{equation}\label{eq:Z}
    Z = Z_{\rm ISM,0} (1 - \ssn - \sicm) + Z_{\rm SN}\ssn + Z_{\rm ICM}\sicm.
\end{equation}

As presented in \citet{2020ApJ...895...43S}, the dominance of the mixing-driven momentum transfer from hot to cool phase can be simply evidenced by the linear relation between the source tracer field and velocity of the cool phase. 
The predicted outflow velocity of the cool phase in the case of mixing-driven momentum transfer is
\begin{equation}\label{eq:vzcool_icm}
    v_{\rm z}^{\rm cool} = \vicm\sicm[cool]
\end{equation}
for ICM-accelerated outflows. In our simulations, both SNe and ICM create the hot gas so that both SN ejecta and ICM tracer fields can leave imprints on the accelerated cool outflows. However, as the relation only holds for the \emph{fresh} tracer field (the tracer field that is first transferred from hot to cool phase), the ICM mass fraction provides an ideal tracer for this purpose, especially for the first acceleration of the ISM. In contrast, as we restarted the simulations from the \noicm{} model after many feedback-star formation cycles, a lot of SN ejecta is already mixed into the cool phase that has been accelerated, fallen back, and reaccelerated many times. The total SN ejecta mass fraction in the cool phase is no longer representative of the amount of the hot gas that is currently mixed. Still, we can see signs of SN accelerated gas from the relatively metal-enriched gas that is moving faster (\autoref{sec:stripping_mixing}).

\section{Overall Evolution} \label{sec:overall}

In this section, we provide an overview of the RPS process in our simulations and visual impressions using a variety of quantities at different times in two representative models.

\subsection{Overview of RPS in Simulations}\label{sec:overview}

\begin{figure*}
    \centering
    \includegraphics[width=0.49\textwidth]{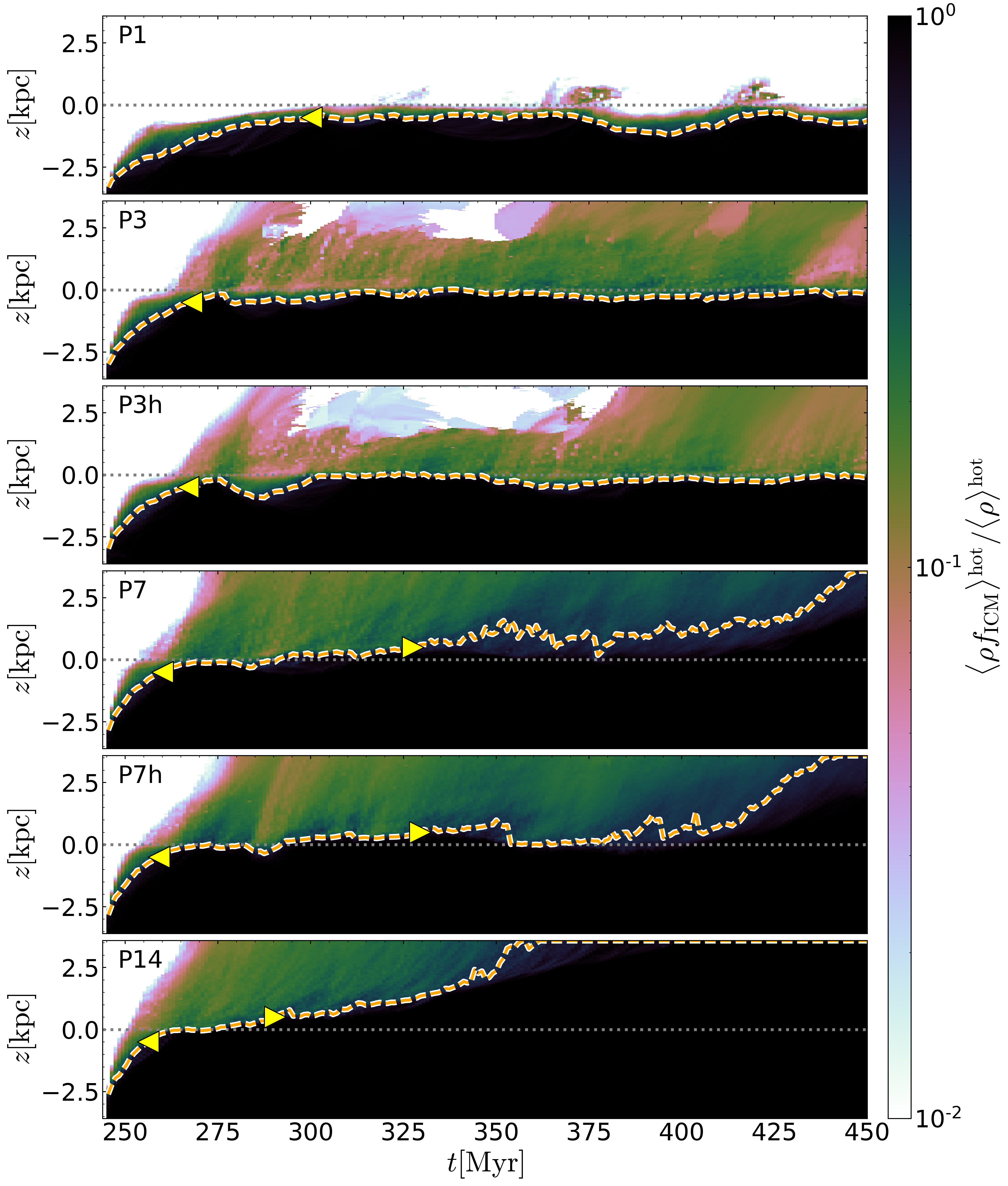}
    \includegraphics[width=0.49\textwidth]{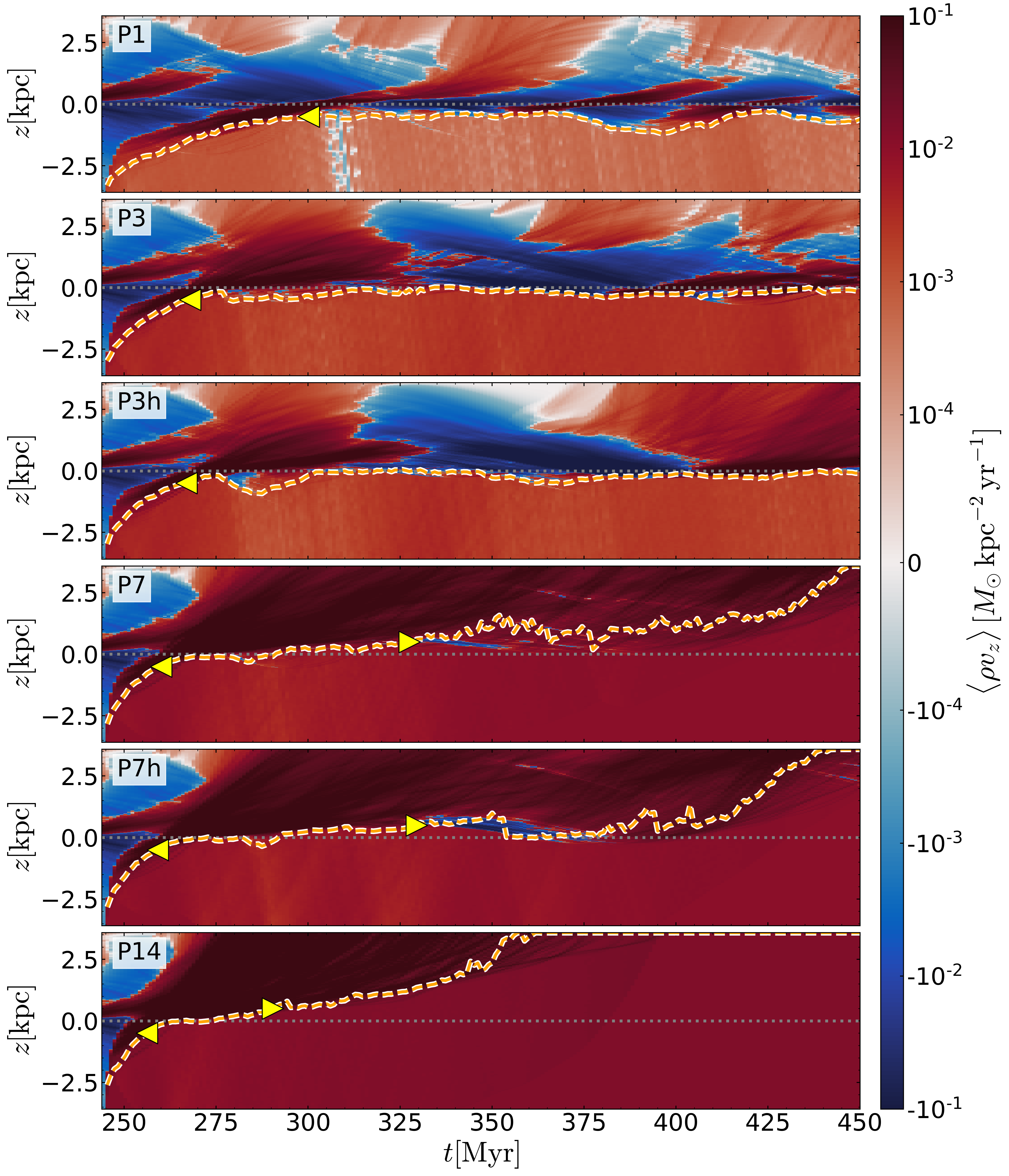}
    \caption{Horizontally-averaged ICM mass fraction in the hot phase (left) and vertical momentum density (= upward mass flux; right) as a function of time for all ICM models. The symmetric plane of the external gravity ($z=0$) is indicated by the horizontal dotted lines. The orange dashed line denotes the ICM-ISM interface as defined by $\overline{\sicm[hot]}=0.5$. Two left- and right-pointing triangles demark the beginning of the early and active stripping stages as defined by the earliest time at which the ICM-ISM interface reaches $z=-500\pc$ and $500\pc$, respectively.}
    \label{fig:sicm}
\end{figure*}

To summarize the general response of the ISM as a whole to the ICM inflows, \autoref{fig:sicm} plots the vertical profiles of the ICM mass fraction in the hot phase (left) and vertical momentum density (= upward mass flux; right) as a function of time for all ICM models. The former is defined by the mass-weighted horizontal average of $\sicm$ of the hot gas, $\overline{\sicm[hot]}\equiv\abrackets{\rho\sicm}^{\rm hot}/\abrackets{\rho}^{\rm hot}$. From top to bottom, we show all models in ascending order of the ICM ram pressure including two high-resolution models shown in the 3rd and 5th rows. We plot the reference line of $\overline{\sicm[hot]}=0.5$ that defines the mean boundary of the ICM-ISM. Owing to the multiphase structure of the ISM, actual boundaries between the ICM and ISM are much more complex (see \autoref{sec:morphology}). The positive (upward) mass flux below the interface simply represents the mass flux of the ICM inflows. As soon as the simulations restarted with the ICM inflows, the ISM in the bottom half is quickly pushed up, and the ICM-ISM interface approaches the midplane in 10-50~Myr depending on the ICM inflow strength. Then, the interface either remains near the midplane with clear separation in the ICM fraction (weak ICM models) or continuously marches upward (strong ICM models). This dichotomy is in excellent agreement with the expectation based on the simple stripping condition listed in Column (5) of \autoref{tbl:models}. 

Using the position of the ICM-ISM interface, we divide overall evolution into three stages; compression stage, early stripping stage, and active stripping stage. The earliest time at which the ICM-ISM interface reaches $z=-500\pc$ and $z=500\pc$ respectively defines the beginning of the early and active stripping stages (marked by left- and right-pointing triangles in \autoref{fig:sicm}).

Because the multiphase ISM is porous and has low-density channels through which the ICM can penetrate, the ICM gradually pollutes the ISM in the upper disk even when the interface stays near the disk midplane. The only exception is the \icmpww{} model, where the penetration of the ICM is not effective, and the ISM in $z>0$ remains unpolluted over the entire simulation duration ($\sicm<1\%$). As a result, the mass flux evolution in the upper disk of the \icmpww{} model is qualitatively similar to that in the \noicm{} model (\autoref{fig:noICM}(b)), indicating that the outflows are still driven by SN feedback. In the \icmpw{} model, the ICM mass fraction in the upper half increases quickly and becomes larger than 10\%. The mass flux in this model is overall enhanced, while the fountain component (alternating positive and negative signs) still exists, implying insufficient acceleration of the cool phase with the marginally weak ICM. The high-resolution model (\icmpwh{}) behaves essentially the same, while the late time evolution shows a strong outflowing epoch. Given the inherently stochastic nature of the evolution, this difference should not be interpreted as systematic resolution dependence. Rather, the qualitative similarity of the early evolution ($t<350\Myr$) means the convergence of the overall evolution.

The ICM penetration in both \icmps{} and \icmpss{} models is highly efficient and leads to the immediate enhancement of the ICM mass fraction in the upper disk. The ICM-ISM interface continuously moves upward in the \icmpss{} model, but the \icmps{} model spends a quite long time ($\sim 70\Myr$) in the early stripping stage with a significantly larger ICM mass fraction ($>10\%$) in the upper disk than that in the \icmpw{} model. The net mass flux in the upper disk is always positive in the strong ICM models, demonstrating the dominant role of the ICM in driving outflows and implying RPS in action. Again, the high-resolution model (\icmpsh{}) shows a very similar evolution with its low-resolution counterpart.

We emphasize that the multiphase RPS occurs continuously in both early and active stripping stages for the strong ICM models. In the early stripping stage, the ICM finds low-density channels in the porous ISM to penetrate. In doing so, the ICM begins to shred the ISM and transfer mass, momentum, and energy while mixing occurs. In the active stripping stage, which only exists in the strong ICM models, the ICM fills the volume in a wide range of the disk. The ICM mixing and momentum transfer occur almost all around the simulation volume, and the ISM is effectively accelerated and removed from the simulation domain. We will delineate the mixing-driven stripping in \autoref{sec:stripping}.

\subsection{Time Evolution of Masses}\label{sec:tevol}

\begin{figure}
\centering
\includegraphics[width=1\linewidth]{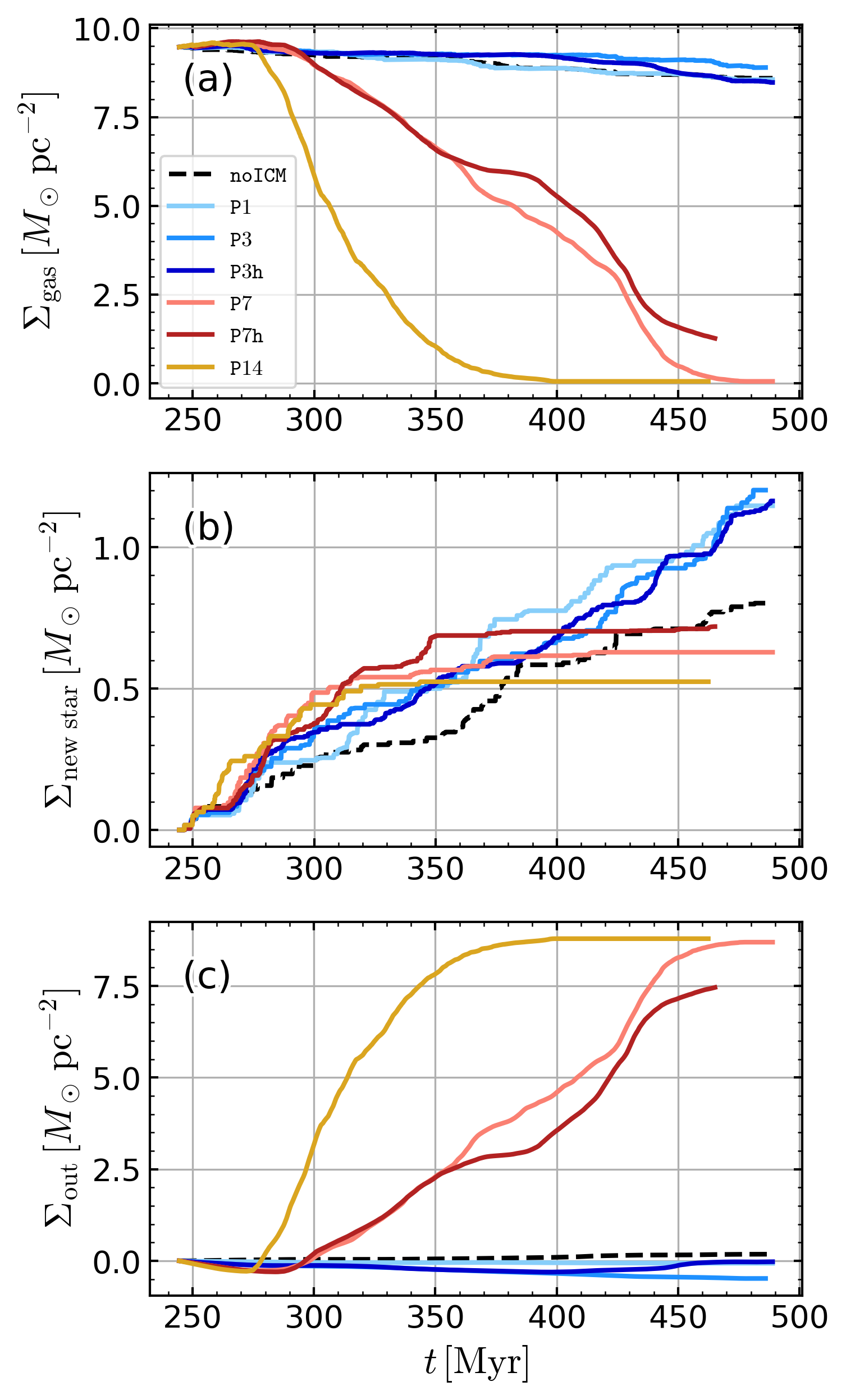}
\caption{Time evolution of (a) total (ICM+ISM) gas surface density, (b) stellar surface density of newly formed stars, and (c) surface density of outflowing gas. (b) and (c) are cumulatively calculated by counting all new stars' mass and integrating mass fluxes at both upper and lower boundaries from the restart of the simulations. The colored solid lines correspond to the models with the different ICM pressure, while the black dashed line is for the \noicm{} model.}
\label{fig:surf}
\end{figure}

\autoref{fig:surf} shows the time evolution of (a) total (ICM+ISM) gas surface density, (b) surface density of new stars $\Sigma_{\rm new-star}$, and (c) surface density of gas passed through the vertical boundaries $\Sigma_{\rm out}$. $\Sigma_{\rm new-star}$ is defined by summing up the total mass of stars formed since we restart the simulations, and $\Sigma_{\rm out}$ is calculated by integrating the net mass flux at the vertical boundaries (both escaped to the top and injected from the bottom) over time. A clear dichotomy between the weak and strong ICM models is visible. The strong ICM models lose gas and stop forming stars after $\sim 100\Myr$. The weak ICM models retain (or even gain) gas and form more stars than the \noicm{} model.
Despite the highly complex interaction between the multiphase ISM and ICM revealed in our simulations (as discussed in \autoref{sec:morphology}), the simple stripping condition estimated by \autoref{eq:Wgg} provides a reliable prediction for the fate of gas disk. This is in part because we only model the face-on, plane-parallel interaction, ideal for the simple criteria to work best.

Even without the ICM, the \noicm{} model also loses its gas through star formation and outflows powered by SN feedback \citep{2018ApJ...853..173K,2020ApJ...900...61K}. The mean SFR and mass outflow rate over the simulation duration of $250-500\Myr$ are $\Ssfr = 3.1\times10^{-3}\sfrunit$ and $\dot{\Sigma}_{\rm gas,out} = 7.7\times10^{-4} \sfrunit$, respectively.
With the ICM inflows, the total gas mass within the domain can increase as the ICM is added to the system unless the outflow and SFRs are greatly increased. The \icmpww{} model closely follows the evolution curve of the \noicm{} model as the enhancement in SFRs is compensated by the decrease in outflow rates (panel (a)). In the \icmpw{} model, $\Sigma_{\rm out}$ becomes negative, implying that the net inflow through the boundaries (panel (c)). Overall gas mass still decreases when taking into account the loss due to star formation. In the \icmps{} and \icmpss{} models, the gas mass decreases quickly as the outward mass fluxes are greatly enhanced after the compression stage (see also \autoref{fig:sicm}). The gas compression by the ICM causes enhancement of the early star formation in all ICM models (see \autoref{sec:sfr} for in-depth analysis). The half-mass stripping time defined by the time interval between $\Sigma_{\rm gas}(t)=\Sigma_{\rm gas,max}$ and $\Sigma_{\rm gas,max}/2$ is $\sim 130\Myr$ and $60\Myr$ for the \icmps{} and \icmpss{} models, respectively. At around similar time scale, star formation in the \icmps{} and \icmpss{} models is completely quenched, showing a flattening in $\Sigma_{\rm new-star}$ (panel (b)). $\Sigma_{\rm out}$ flattens later after complete stripping (the ICM flows freely; panel (c)).

Overall qualitative behaviors are converged with the resolution, but the later time evolution shows differences mainly due to the stochasticity of the evolution.

\subsection{Morphological Evolution} \label{sec:morphology}

\begin{figure*}
    \centering
    \includegraphics[width=\textwidth]{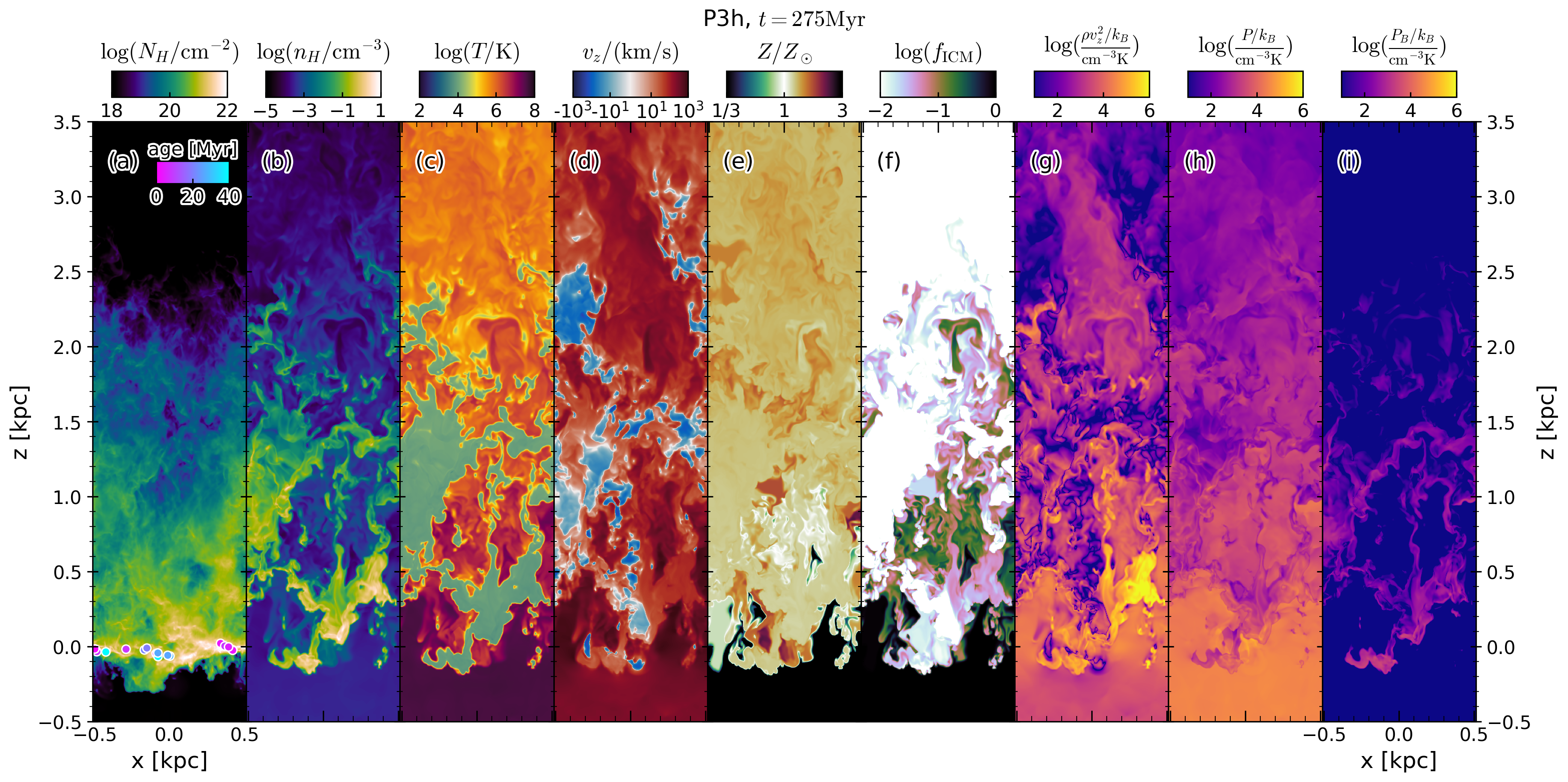}
    \includegraphics[width=\textwidth]{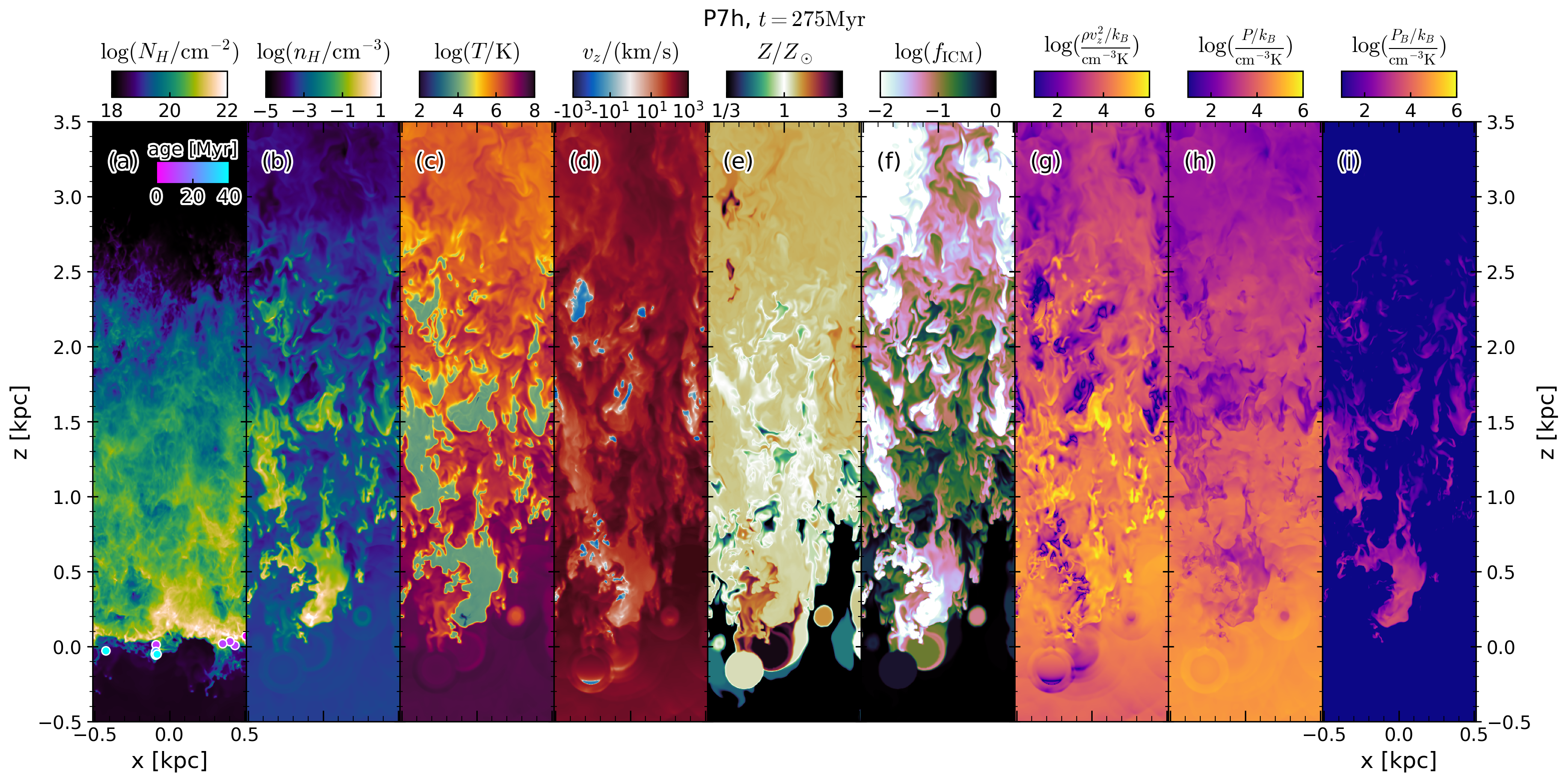}
    \caption{Detailed visualization of the multiphase ISM interacting with the ICM at the end of the compression stage ($t=275\Myr$) for the \icmpwh{} (top) and \icmpsh{} (bottom) models. In (a), column density integrated along the $y$-axis is shown with sink particles colored by their age. The other columns show physical quantities in one-zone thick slice centered at $x=0$. From left to right, we show (b) hydrogen number density ($n_H$), (c) temperature ($T$), (d) vertical velocity ($v_z$), (e) metallicity ($Z$), (f) ICM mass fraction ($\sicm$), (g) ram pressure ($\rho v_z^2$), (h) thermal pressure ($P$), and (i) magnetic pressure ($P_B = B^2/8\pi$). Only $z>-0.5\kpc$ region is shown to focus on the upper disk. Animations of this figure for each model are available in the electronic journal. The video begins at $t=244\Myr$ and ends at $t=450\Myr$. The real-time duration of the video is 20 s.} 
\label{fig:slc_early}
\end{figure*}

\begin{figure*}
    \centering
    \includegraphics[width=\textwidth]{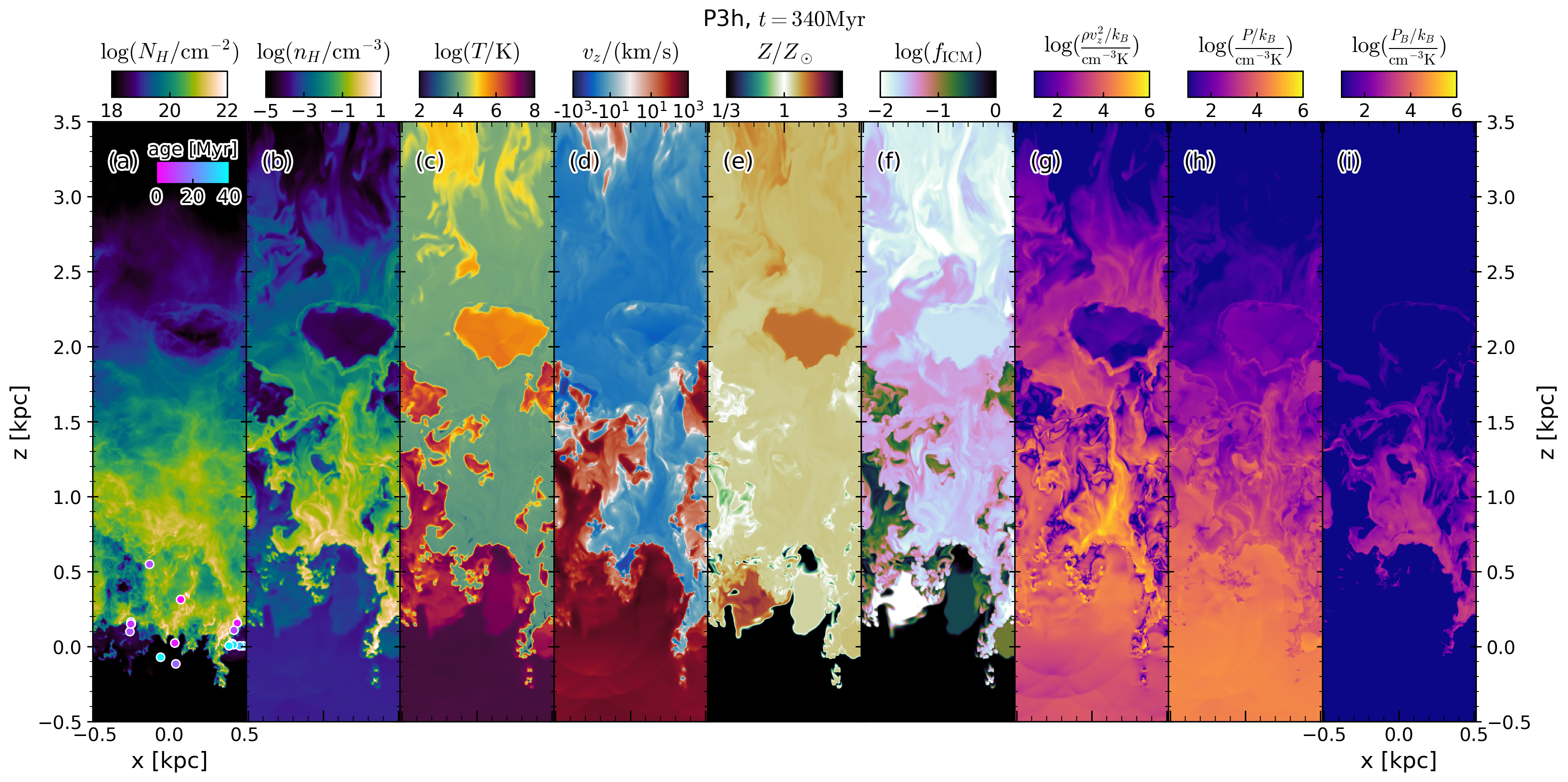}
    \includegraphics[width=\textwidth]{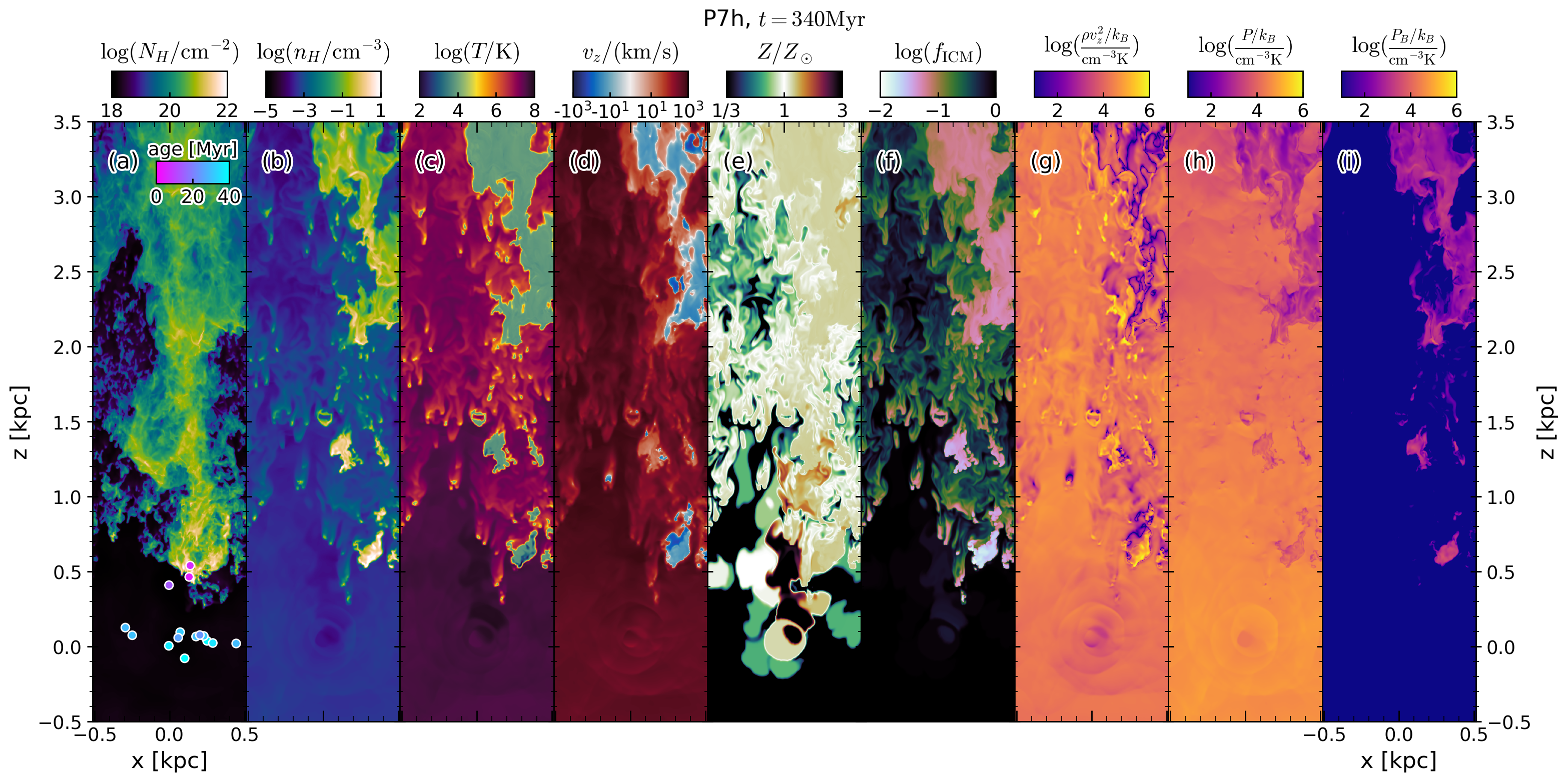}
\caption{Same as \autoref{fig:slc_early}, but at $t=340\Myr$.}
\label{fig:slc_late}
\end{figure*}

To delineate the interaction between the ICM inflows and the multiphase ISM, \autoref{fig:slc_early} shows snapshots at $t=275\Myr$ for two representative models, \icmpwh{} (top) and  \icmpsh{} (bottom). At this epoch, the ISM in the bottom half has already pushed up the midplane in both models, so we only show the upper disk $z>-0.5\kpc$. Visual impressions between projected density and slices are substantially different. The projected density shows an overall density distribution with mild fluctuations, with similar sharp cutoffs at around $z=0$ for both models. The immediate impression might be that there is a well-defined ICM-ISM interface near the midplane. However, the slices unveil a highly-porous, multiphase structure with a large density and temperature contrast. Also, significant penetration of the ICM (see (f)) depicts a significant difference between the two models. We emphasize that, compared to the projection maps, slices (or thin projections) of gas physical quantities are often more useful to deliver visual insights. This is particularly true when looking into the interaction between the different phases with large contrasts in physical properties.

The density (panel (b)) and temperature (panel (c)) slices of both models show a large, cool gas structure that begins to face the ICM near the midplane. This structure looks like a continuous, single structure in density and temperature, but $v_z$ (panel (d)) and hence $\rho v_z^2$ (panel (g)) show a sharp change between the left and right sides. The enhanced ICM mass fraction $\sicm$ (panel (f)) and the relatively large outward velocity (panel (d)) in the right side of this structure imply that this is the gas originally in the bottom half accelerated by the ICM. The accelerated cool gas from the lower disk remains intact and appears as a high ram pressure chunk at around $z=0.5\kpc$ in the \icmpwh{} model, but it is already substantially shredded and fragmented in the upper disk in the \icmpsh{} model. Another interesting cool gas structure that shows a difference between the two models is one located at $z\sim 1\kpc$ near the left edge of the slice. In the \icmpwh{} model, this structure is still falling (panel (d)), while the same structure has already significantly shredded and accelerated by the interaction with the ICM in the \icmpsh{} model. There are plenty of similar falling gases (fountain flows) in the \icmpwh{} model across a wide range of $z$. In the \icmpsh{} model, such infalling fountain flows are no longer prevalent, but they are generally more compressed and even outflowing.

Generally, the ICM in the \icmpsh{} model manages to intrude almost the entire regions of the upper disk. It is visually evident from the enhanced $\sicm$ seen in most regions, which correspond to the enhanced ram and thermal pressure and outward velocity. The metallicity (panel (e)) contains more complex information due to the additional contribution of the high-metallicity SN ejecta. As noted in \autoref{sec:method_phase}, the ISM has been enriched in the \noicm{} model over $\sim250\Myr$. The mean ISM metallicity at the beginning of the ICM models (shown as bright orange color in panel (e)) is thus larger than the initial metallicity $Z_\odot$. Without the ICM, it is the high metallicity gas injected by SNe that is filling low-density regions (it is still the case for $z>2\kpc$ in the \icmpwh{} model). With the ICM, now the low metallicity gas is filling the low-density regions (more evident in the \icmpsh{} model), which also show high pressures (panels (g) and (h)). As the high-pressure ICM compresses the ISM, the magnetic pressure is enhanced in cooler, denser gas (panel (i)). 

In \autoref{fig:slc_late}, we select snapshots at $t=340\Myr$ for the \icmpwh{} (top) and \icmpsh{} (bottom) models. The late-time evolution differs in the two standards more substantially and is even evident in the density projection (panel (a)). Since the ICM inflow alone cannot keep pushing the ISM away in the \icmpwh{} model, the bulk ISM is falling back as star formation has suppressed. In contrast, the strong ICM inflow alone can continue to strip the ISM in the \icmpsh{} model (see also \autoref{fig:sicm}). While disturbed significantly, in the \icmpwh{} model, the ICM fraction in the cool phase is still less than a few percent, and overall visual impression is not very different from what is shown in \autoref{fig:slc_early}. The hot ICM keeps penetrating through the low-density channels, creating shearing interfaces between the cool and hot phases in which the majority of mixing occurs. The volume fractions of the hot and cool phases are comparable all over the upper disk. We note that when the hot gas is only created by SNe in the \noicm{} model, the hot gas volume fraction is $\sim20-30\%$ near the midplane and increases to $\sim 50\%$ at $z\sim1\kpc$ and to $\sim 100\%$ at $z>2\kpc$ \citep{2020ApJ...897..143K}. Star formation continues in this model at slightly higher rates than the \noicm{} model. Once stars form, they fall faster than gas as stars do not feel the ICM pressure. This causes SN feedback in the ICM dominated regions below the midplane, sometimes creating metal-enriched hot bubbles between the hot ICM and cool ISM (panels (e) and (f)). 

At this time, the \icmpsh{} model already loses about $\sim 30\%$ of its total mass, and almost all ISM is pushed above $z\sim0.5\kpc$ (panel (a)). The cool ISM is highly fragmented and confined in smaller volumes (panels (a) and (b)). The dense gas structure facing the ICM at $z\sim0.5\kpc$ is the leftover from the first major stripping of the main ISM disk and falling. Stars were just born in this strongly compressed structure, which is the final major star formation event before complete quenching (there will be additional extraplanar star formation in the structure far above the midplane later). Star clusters formed at high $z$ during the previous evolutionary stage have fallen below the ICM-ISM interface, some of which still host SNe and create metal enriched bubbles at $z\sim0\kpc$. Even in the cool gas above $z>2\kpc$, the ICM fraction is quite high $\sim 10\%$ as the ICM is continuously mixed in to transfer momentum and maintain momentum flux against the weight at that height (\autoref{sec:stripping_mixing}). The acceleration was not sufficient to blow away this structure though. Except for these distinctive large structures, smaller clouds have already been ablated as evidenced by many tadpole structures whose density and temperature are respectively higher and lower than those of the typical hot ICM. The intermediate phase gas populated in the wakes of the front clouds in part escapes the domain and in part condenses back to the cool phase especially when the wakes meet the large cool structure in the back. From panels (d) to (g), it is evident that such acceleration is most efficient in the envelope of the large cool phase structure, where both ICM fraction and vertical velocity are relatively high. 

\section{Stripping of the multiphase ISM}\label{sec:stripping}

The acceleration and stripping of the ISM is a generic feature of the disk interacting with the ICM. To provide a more quantitative view, we first investigate when and where the hot ICM exchanges its mass, momentum, and energy with the cool ISM. We then seek evidence of mixing-driven momentum transfer.

\subsection{Mass, Momentum, and Energy Transfers between Thermal Phases}\label{sec:stripping_transfer}

In this subsection, we define physical quantities averaged over a range of volume and time to understand how different gas phases exchange their mass, momentum, and energy at different regions and times. We begin by integrating the conserved form of the MHD equations over the entire horizontal area $A=L_xL_y$ and chosen vertical range ($z\in(\zmin,\zmax)$) to obtain a set of conservation equations
\begin{equation}\label{eq:qdot}
    \dot {q} + [\mathcal{F}_q(z_{\rm max}) - \mathcal{F}_q(z_{\rm min})] A = +\dot{q}_{\rm source} - \dot{q}_{\rm sink},
\end{equation}
where $q=M$, $p$, and $E$ for mass, vertical momentum, and total energy, which are respectively defined as
\begin{equation}\label{eq:mass_mom}
    M \equiv \int_{\zmin}^{\zmax} \rho dV, \textrm{and}\quad p \equiv \int_{\zmin}^{\zmax} \rho v_z dV,
\end{equation}
and
\begin{equation}\label{eq:energy}
    E \equiv \int_{\zmin}^{\zmax} \rbrackets{\frac{1}{2}\rho v^2 + \frac{P}{\gamma-1} + P_B} dV.
\end{equation}
The horizontally-averaged fluxes (the square bracket term in the left lend side of \autoref{eq:qdot}) are respectively defined by for mass, vertical momentum, and total energy as
\begin{equation}\label{eq:massflux}
    \mathcal{F}_M(z) \equiv \frac{1}{A}\int \rho v_z dx dy,
\end{equation}
\begin{equation}\label{eq:momflux}
    \mathcal{F}_p(z) \equiv \frac{1}{A}\int \rbrackets{\rho v_z^2 + P + P_B- \frac{B_z^2}{4\pi}} dx dy,
\end{equation}
and
\begin{equation}\label{eq:eneflux}
    \mathcal{F}_E(z) \equiv \frac{1}{A}\int \rbrackets{\rho v_z \rbrackets{\frac{v^2}{2} + \frac{\gamma}{\gamma-1}\frac{P}{\rho}} + \mathcal{S}_z} dx dy,
\end{equation}
where the adiabatic index $\gamma=5/3$ and the vertical component of the Poynting flux is $\mathcal{S}_z\equiv  (v_z B^2 - B_z \vel\cdot\mathbf{B})/(4\pi)$. Note that we do not include the gravitational potential term in the energy flux such that the work done by gravity is appeared as a sink term in the right hand side of \autoref{eq:qdot}, similarly to the momentum sink term by weight.

In the mass conservation equation, we have mass sink due to star formation ($\dot M_*$) and mass source due to SN ejecta ($\dot M_{\rm SN} = \dot{N}_{\rm SN}M_{\rm ej}$) if stars are born and SNe are exploded in the particular volume of interest. For our chosen stellar population synthesis model (STARBURST99 with the Kroupa IMF; \citealt{1999ApJS..123....3L}), 1 SN ejects $10\Msun$ for $\sim100\Msun$ of new star formed, implying $\dot{M}_{\rm SN} \sim 0.1 \dot{M}_*$ on average. We note that we did not subtract the mass of stars exploding as SNe in the simulations. The gravitational weight acts as sink of vertical momentum, $\dot{p}_{\rm sink} = \mathcal{W} A$ and
\begin{equation}\label{eq:weight}
    \mathcal{W} \equiv \frac{1}{A} \int_{\zmin}^{\zmax} \rho\frac{d \Phi}{dz} dV,
\end{equation}
where $\Phi = \Phi_{\rm ext} + \Phi_{\rm sg} + \Phi_{\rm tidal}$ includes the external potential (\autoref{eq:phi_ext}), the self-gravity potential returned from the solution of the Poisson's equation for both gas and star cluster particles, and the tidal potential arising from the local rotating frame $\Phi_{\rm tidal} = - q\Omega^2 x^2$.
Finally, there are energy sources from SN energy injection and shearing-box stress $\dot{E}_{\rm source} = \dot{N}_{\rm SN} E_{\rm SN} + w_{\rm xy}$, where $w_{\rm xy} = (q\Omega L_x)\int [\rho v_x\delta v_y - B_x B_y/(4\pi)]dydz$ is integrated over either of the $x$ boundaries \citep[][]{1995ApJ...440..742H}, and sinks from net radiative cooling $\dot{E}_{\rm cool} = \int (n_H^2\Lambda(T) - n_H\Gamma)dV$ and the work done by gravity
\begin{equation}\label{eq:grav_work}
\dot{E}_\Phi \equiv \int_{\zmin}^{\zmax} \rho \vel\cdot\nabla \Phi dV.
\end{equation}

As we exclusively assign the gas in different thermal phases, everything is separable by thermal phases. Since stars are formed from the cool (more precisely, cold) gas, mass sink by star formation can be accounted for the cool phase. We include SN mass and energy source to the hot phase; only a small fraction ($\lesssim 10\%$) of unresolved SNe injects mass and energy into the form of the cooler phases. By integrating \autoref{eq:qdot} over a time interval $\Delta t$, we obtain
\begin{equation}\label{eq:qdot2}
\sum_{\rm ph}\dot{q}_{\rm net}^{\rm ph} = 0,
\end{equation}
where
\begin{equation}\label{eq:qnet}
\dot{q}_{\rm net}^{\rm ph} \equiv \frac{\Delta q^{\rm ph}-\Delta q_{\rm source}^{\rm ph}+\Delta q_{\rm sink}^{\rm ph}}{\Delta t} + \sbrackets{F_{q, u}^{\rm ph} - F_{q, l}^{\rm ph}}A,
\end{equation}
where ph = cool, int, and hot (see \autoref{tbl:phase}).
Here, $\Delta q$ means the temporal difference of mass, vertical momentum, and total energy. Similarly, the temporal difference of cumulative mass and energy injection by SNe and stellar mass formed defines SN source and star formation sink. The weight and gravity work terms are defined by time integration. 
The net cooling term is calculated by time integration of the instantaneous net cooling rate from simulation outputs.
Similarly, the time-averaged fluxes at upper and lower faces are defined by
\begin{equation}\label{eq:Favg}
    F_{q, u/l} \equiv \frac{1}{\Delta t}\int_{t}^{t+\Delta t}  \mathcal{F}_q(z_{\rm max/min}) dt.
\end{equation}
By fully accounting for the temporal changes of each quantity, fluxes through vertical faces, and source/sink in each phase, $\dot{q}_{\rm net}^{\rm ph}$ represents any loss/gain of mass, momentum, and energy through phase transitions within space-time bins of interest.

In practice, we measure each term using output snapshots dumped every $\sim1\Myr$. The cadence of snapshots was not fine enough to satisfy the conservation perfectly (\autoref{eq:qdot2}) as this \textit{post-processing} assumes that each variable involved in the time integration (e.g., cooling and flux terms) is constant over the snapshot interval of $1\Myr$.\footnote{One can use the \emph{instantaneous} conservation equations (\autoref{eq:qdot}). In this case, however, time derivative terms can be noisy and inaccurate.} Bearing this caveat in mind, we analyze mass, momentum, and energy transfers between phases in space and time bins and only consider them to be reliable when the net changes are distinctive over the level of non-conservation errors.

\subsubsection{Mass Transfer}\label{sec:phase_transition}

\begin{figure*}
    \centering
    \includegraphics[width=\textwidth]{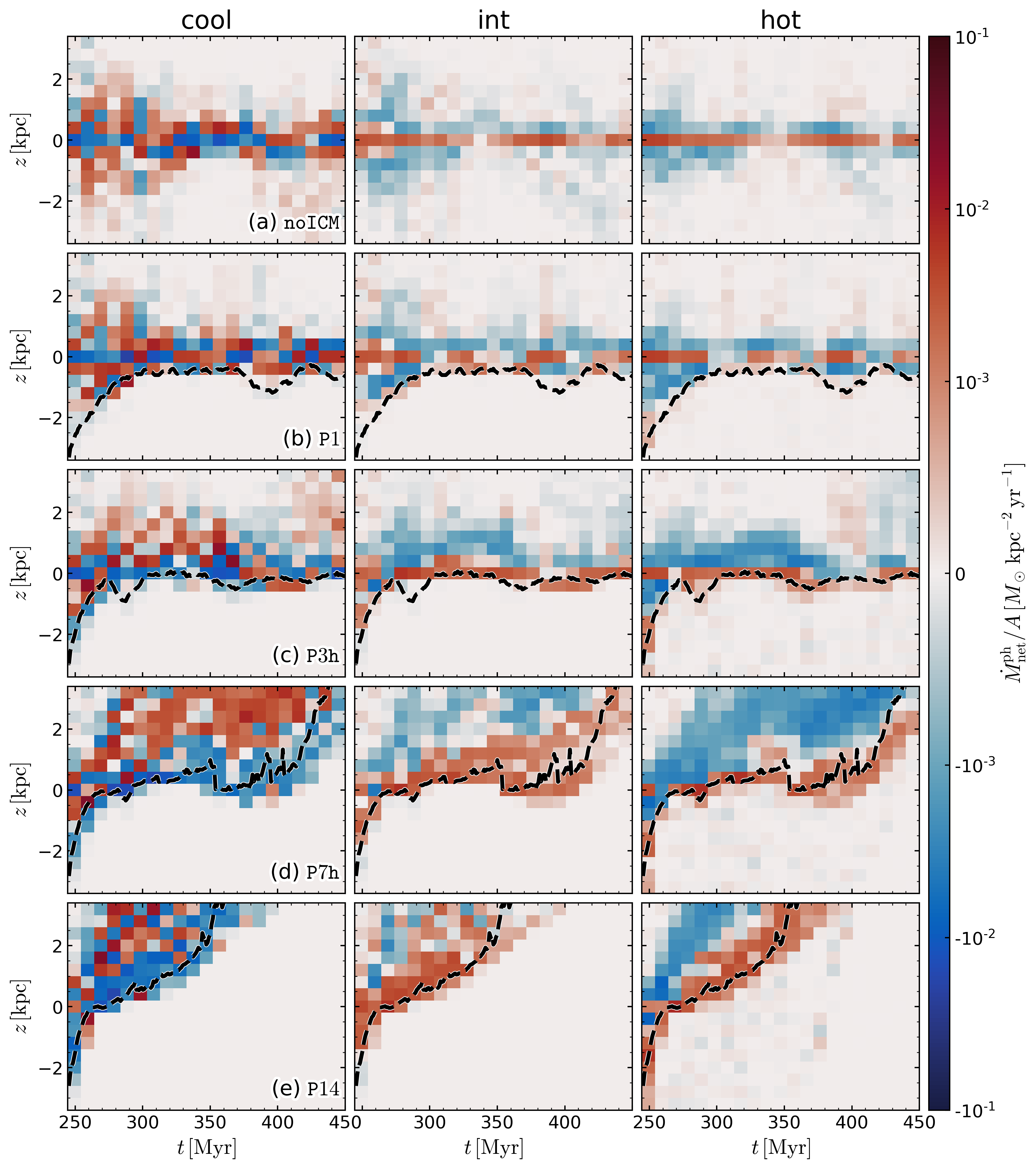}
    \caption{Net mass change rates per unit area for each phase due to phase transition. The space-time bin is separated by $\Delta t=10\Myr$ and $\Delta z=400\pc$. Mass sink (source) by star formation (SN ejecta) is calculated in each space-time bin and added (subtracted) to the cool (hot) phase to isolate the gain and loss solely due to phase transition. The dashed lines in each panel show the ICM-ISM interface as defined in \autoref{fig:sicm}. The phase transition driven by the ICM shows a general trend (clearer with stronger ICM pressure) described by the gain (loss) of the hot (cool) phase near the ICM-ISM interface followed by the loss (gain) of the cool (hot) phase in the extraplanar region.
    }
    \label{fig:mdot_tz}
\end{figure*}

We first consider mass conservation to understand phase transition between thermal phases. \autoref{fig:mdot_tz} plots the net mass gain (red) and loss (blue) rates of cool, intermediate, and hot phases from left to right within the space-time bins. Here, we consider $\Delta z = 400\pc$ thickness slabs centered at $z = -3.6,\, -3.2,\, \cdots, -0.4,\, 0,\, 0.4\, \cdots,\, 3.2,\, 3.6\kpc$ and time interval of $\Delta t =10\Myr$. The mass sink by star formation is added in the cool phase (left column) using new star particles formed in given space-time bins. Similarly, the mass source by SN ejecta is subtracted in the hot phase (right column). The positive (red) and negative (blue) values in \autoref{fig:mdot_tz} are solely due to phase transition within the space-time bins.

Without the ICM inflows (\noicm{}; row (a)), the net gain of the hot and intermediate phases at the midplane stands out as thin red strips. In this case, it is clustered SNe that creates the hot phase via shock-heating of the ambient medium, showing the net loss (blue strip) in the cool phase. In our simulations, superbubbles expand into an inhomogeneous ambient medium, also creating a lot of the intermediate phase through the ablation of the cool phase shown as the net gain in the intermediate phase.

Both hot and intermediate phases subsequently cool above and below the midplane slab. Sometimes, this ``cooling'' region is further extended to $z\sim\kpc$. It can be both direct cooling of the hot shocked gas (shell formation; e.g., \citealt{2015ApJ...802...99K}) or mixing of the hot and cool gas (interface mixing; e.g., \citealt{2017ApJ...834...25K,2019MNRAS.490.1961E}). If the hot bubbles have completely cooled within the thickness of the midplane slab we consider ($z=\pm200\pc$), the net gain/loss in different phases will not be visible. In other words, most superbubbles in our simulations can expand with their radii larger than $200\pc$ over $10\Myr$ before they cool.

The upper half of the \icmpww{} model shows overall similar results with the \noicm{} model. As the ICM ram pressure gets stronger, noticeable differences begin to appear. In the \icmpwh{} model, the layer in which the hot and intermediate phases gain the mass remains thin, while the cooling region is more extended toward higher-$z$ and becomes more prominent. The ICM-ISM interface stays near the midplane, and the effect of the ICM appears as the enhanced gain of the hot (and intermediate) phase in the midplane slab as the hot ICM shocks the cool ISM. At $t\sim 400\Myr$, the net gain in the hot phase is visible beyond the midplane, which is due to the successful penetration of the hot ICM through the entire upper disk (see \autoref{fig:sicm}). This breakout causes cool-to-hot phase transition followed by hot-to-cool phase transition at $t\sim 420-450\Myr$.

With even stronger ICM inflows, the mass gaining layer for the hot and intermediate phases gets thicker (slightly thicker for the intermediate phase). Now, the cool-to-hot phase transition is dominated by the interaction between the hot ICM and the cool ISM rather than SN feedback, especially at late times when star formation is nearly quenched. The large energy flux carried by the hot ICM can heat and ablate a significant amount of the cool phase, converting the cool ISM into the hot phase while populating the intermediate phase in mixing layers. Above the cool-to-hot phase transition layer, there is an extended region where the hot-to-cool phase transition occurs. In this region, best viewed in \icmpsh{}, the hot and intermediate phases cool back to the cool phase. What is happening here is more like precipitation of the hot (with intermediate) phase into the volume filling cool phase that has been pushed by earlier interaction. This is somewhat different from the cooling of the mixed gas itself that drives phase transition from hot to cool in the cloud wakes as seen in radiative cloud-crushing simulations with large sizes \citep[e.g.,][]{2016MNRAS.462.4157A,2018MNRAS.480L.111G,2020MNRAS.492.1970G,2020MNRAS.499.4261S,2021MNRAS.501.1143K,2020MNRAS.492.1841L,2021arXiv210110344A}.

As RPS is much more efficient in \icmpss{}, the hot-to-cool phase transition layer moves quickly outside the simulation domain. At $t>350\Myr$, only the cool-to-hot phase transition occurs within our simulation domain and the gas escapes mostly in the form of the hot phase (see also \autoref{fig:out_in_strong}). While not followed in our simulations, the hot-to-cool phase transition can still occur far above the disk midplane.

\subsubsection{Momentum Transfer}\label{sec:mom_transfer}

\begin{figure*}
    \centering
    \includegraphics[width=\textwidth]{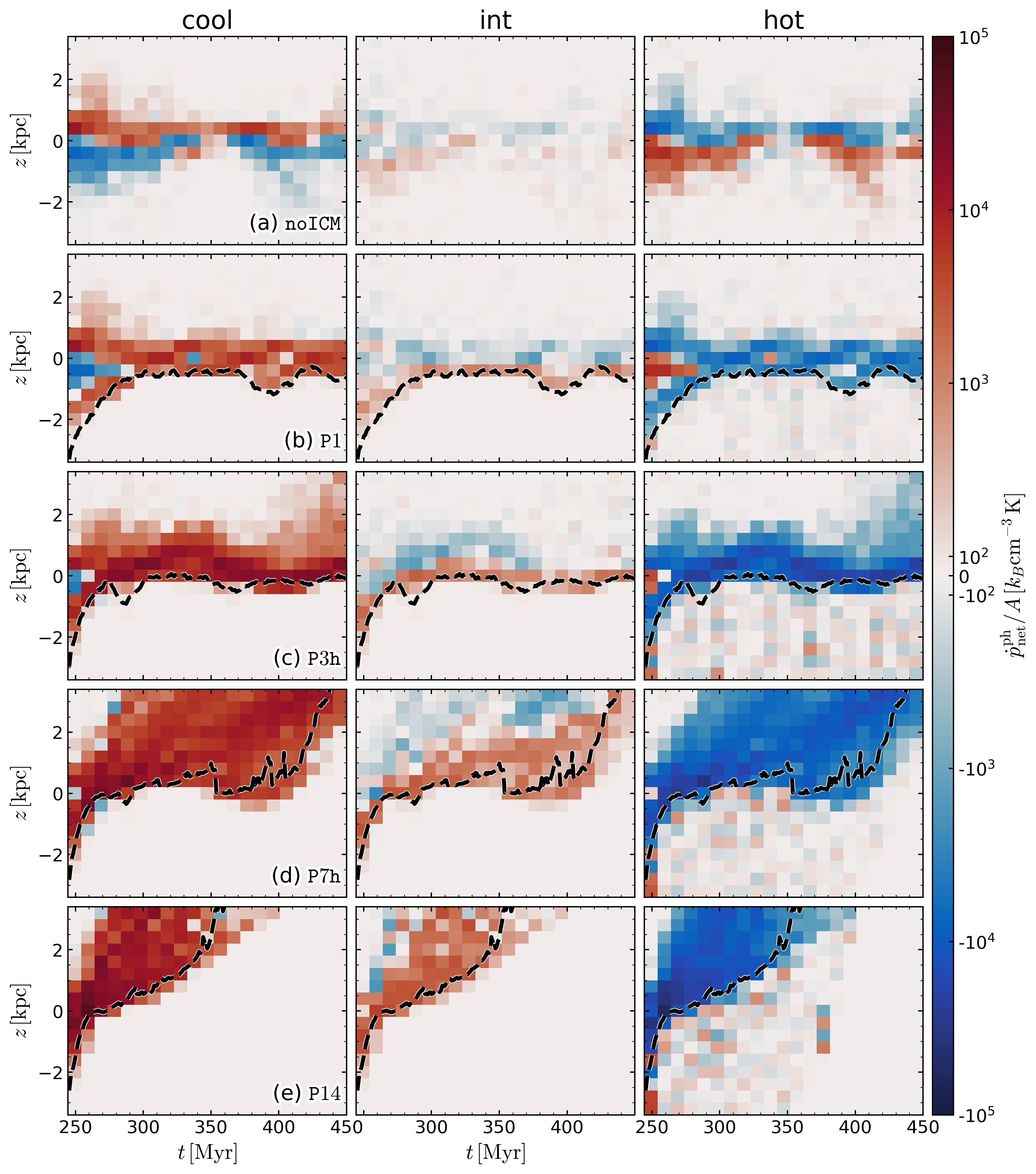}
    \caption{Net vertical momentum change rates per unit area for each phase due to phase transition. The space-time bin is separated by $\Delta t=10\Myr$ and $\Delta z=400\pc$. The weight of gas of each phase (only significant in the cool phase) is included as a sink. The dashed lines in each panel show the ICM-ISM interface as defined in \autoref{fig:sicm}. Noisy, checkerboard-like patterns below the ICM-ISM interface depict the non-conservative errors of the analysis, mainly due to significant time-varying fluxes. The hot ICM's momentum flux is transferred to the cool ISM.}
    \label{fig:pdot_tz}
\end{figure*}

\begin{figure*}
    \centering
    \includegraphics[width=0.9\textwidth]{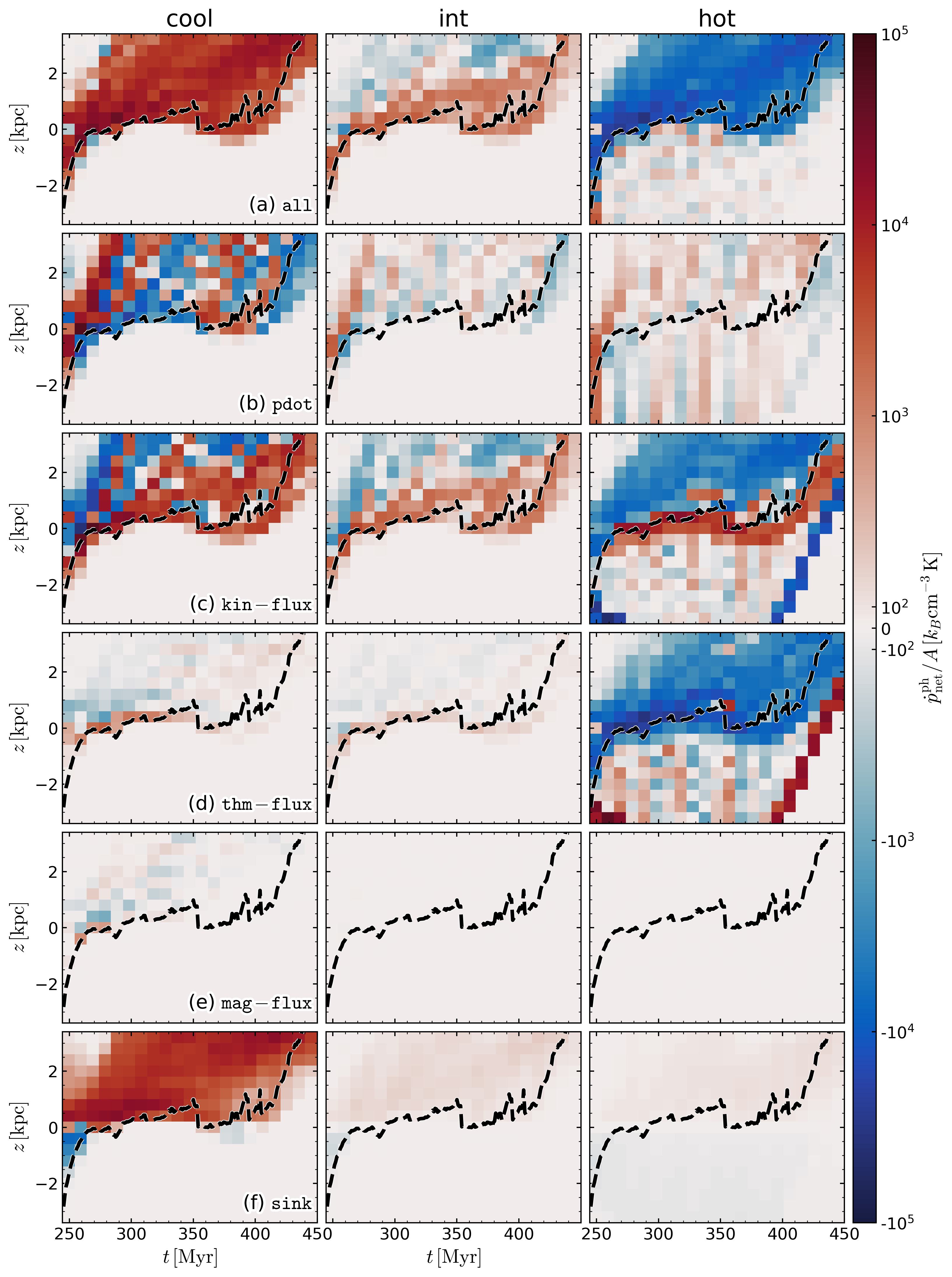}
    \caption{Term-by-term decomposition of vertical momentum change rates (see \autoref{eq:qnet}) for the \icmpsh{} model. We show (a) net momentum change rates per unit area, which is identical to \autoref{fig:pdot_tz}(d). There is no explicit source term, while we show (b) the time dependent term and (f) sink term due to gravity. The flux term is further decomposed into the (c) kinetic, $\rho v_z^2$, (d) thermal, $P$, and (e) magnetic, $P_{\rm mag}-B_z^2/4\pi$, flux terms. The momentum transferred from hot to cool phase is mostly used to provide appropriate support against the increased weight. At later times (active stripping stage), the cool phase gains kinetic flux.
    }
    \label{fig:pdot_tz_P7h}
\end{figure*}

\autoref{fig:pdot_tz} shows net vertical momentum gain/loss rates per unit area for each phase. The weight of each phase is included as a sink, but only the weight of the cool phase is significant. To understand the plot, it is important to keep in mind that the vertical momentum flux $\rho v_z$ has a sign and the positive value (red) means the gain of the upward momentum flux as well as the loss of the downward momentum flux. The \noicm{} model shows the change of sign in each phase about $z=0$. The hot phase loses its upward momentum flux in the upper disk (blue in $z>0$) and downward momentum flux in the lower disk (red in $z<0$) -- in short, the hot phase loses the \emph{outward} momentum flux and the cool phase gains it. In other words, as SN-driven superbubbles expand, the cool phase is accelerated outward by transferring the outward momentum flux of the hot phase. Note that the weight term of the cool phase dominates the net gain (see also \autoref{fig:pdot_tz_P7h}). This means that the continuous momentum transfer from the hot phase (or SN momentum injection) enables the cool phase (mass dominating component) to remain vertically extended (more extended than by the thermal and magnetic support).

In the ICM models, as soon as the ICM-ISM interface reaches the midplane $z=0$, \autoref{fig:pdot_tz} becomes just upward momentum gain/loss in the upper disk. As the ICM pressure gets stronger, the hot ICM dominates the momentum transfer, occurring in a larger region of the upper disk. In the weak ICM models, the momentum transfer is still limited within $z<1-2\kpc$. In these models, the vertical momentum flux gain in the cool phase is simply counterbalanced by the increased weight, confining the disk within the simulation domain (no significant stripping). 

In the \icmpsh{} model, significant hot-to-cool momentum transfers occur all over the upper disk as the ICM fills up a larger volume in this region. In contrast to the mass transfer, which shows a clear dichotomy of the cool-to-hot and hot-to-cool phase transition layers at different heights (\autoref{fig:mdot_tz}(d)), the vertical momentum is always transferred from the hot phase to the cool phase. However, the momentum transfer to the intermediate phase closely follows the mass transfer; it gains momentum when it gains mass. On the one hand, near the ISM-ICM interface, where the mass and momentum transfers have opposite signs in both cool and hot phases, the cool phase is accelerated and gains vertical momentum, while it is shredded and losing mass by hydrodynamic instabilities. Here, the momentum transfer is mainly due to the drag force from the hot gas. The intermediate phase is newly populated by the accelerated and shredded cool phase, gaining both mass and momentum. On the other hand, in the upper region of the disk, the hot phase, together with the intermediate phase, is continuously mixed and cooled back to the cool phase delivering both mass and momentum from hot (and intermediate) to cool phase. There, the momentum transfer is more mixing-dominated.

To help further understanding of vertical momentum transfer in detail, \autoref{fig:pdot_tz_P7h} shows a decomposition of each term in \autoref{eq:qnet} in different phase. From top to bottom, each row shows (a) net gain/loss (identical to \autoref{fig:pdot_tz}(d)), (b) time-dependent term, (c) kinetic, (d) thermal, and (e) magnetic flux terms, and (f) sink term due to weight. In the third column (hot), the ICM shock front marching upward is visible at late times. The hot ICM is thermalized at the shock -- gaining thermal flux and losing kinetic flux. The thermalized hot ICM loses its thermal flux (pressure) due to the interaction with the ISM. This turns into the kinetic flux gains of all phases. The hot kinetic flux gain is due to the mass loading from the cool phase (\autoref{fig:mdot_tz}), and the cool kinetic flux gain represents acceleration by drag force. Note that the thermal flux gains in the cool and intermediate phases are minimal, as the majority of the thermal flux transferred to these phases is radiated away (see \autoref{sec:energy_transfer}). The magnetic flux term is subdominant, while this can be as important as the thermal term in weaker or no ICM models at the midplane. The weight term is dominated by the cool phase as shown in (f), which is larger than the kinetic flux gain of the cool phase at early times. At late times, the weight term becomes comparable to the kinetic flux gain of the cool phase, implying effective stripping. In short, the hot ICM’s momentum flux transferred to the cool phase has been used to extend the ISM vertically and support the increased weight. The kinetic momentum flux of the cool phase shows consistent gains at later times (active stripping phase).

In the \icmpss{} model, where the ICM pressure is so strong that the momentum gain in the cool phase is now dominated by the kinetic flux gain, resulting in actual acceleration of the entire cool phase at a velocity larger than the escape velocity of the simulation domain. The majority of the cool ISM is ablated while it is accelerated and quickly stripped away from the simulation domain.


\begin{figure*}
    \centering
    \includegraphics[width=0.9\textwidth]{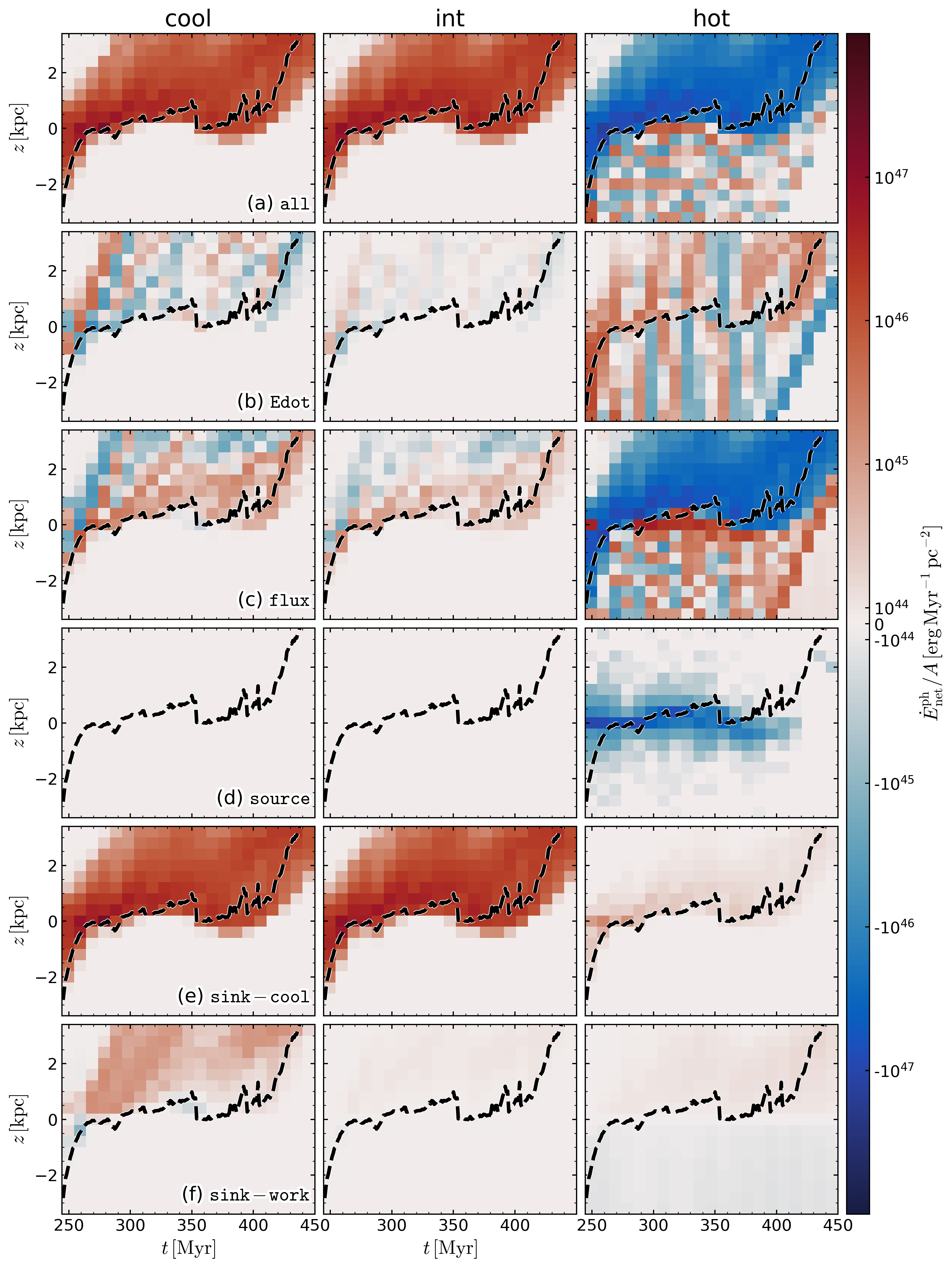}
    \caption{Term-by-term decomposition of energy change rate (see \autoref{eq:qnet}) for the \icmpsh{} model. We show the decomposition of (a) the net energy change rates per unit area
    into (b) the time dependent term, (c) flux term (\autoref{eq:eneflux}), (d) source term (only in the hot phase), (e) sink term by cooling, and (f) sink term by gravity. Noisy, checkerboard like patterns below the ICM-ISM interface depicts the non-conservative errors of the analysis, mainly due to significantly time varying fluxes. The majority of the energy flux transferred from hot to intermediate and cool phases are lost through cooling.}
    \label{fig:Edot_tz_P7h}
\end{figure*}

\subsubsection{Energy Transfer}\label{sec:energy_transfer}

In the simulations, in addition to the energy added by SNe and ICM inflows, there is radiative heating by FUV radiation in the cool phase. But, this radiative heating is balanced by cooling within the same phase. The remaining energy transfer across thermal phases is simple: always from hot to cooler phases regardless of the source of energy in the hot phase. An example, \autoref{fig:Edot_tz_P7h} shows the decomposition of the energy transfer terms for the \icmpsh{} model. The sink term due to radiative cooling in (e) is much larger than the residual energy flux, which is then shared by the actual kinetic flux gain in (c) and work done against gravity in (f) of cooler phases. As seen in \autoref{fig:pdot_tz_P7h}, there is little energy that ends up arriving in the thermal energy/pressure. In this particular model, the source term due to SNe is only significant at early times. Immediately after the ICM-ISM interface reaches the midplane, the overall energy loss is much larger than the explicit source term by SNe (panel (a) and (d)), implying that it is the energy mostly delivered by the ICM inflows (panel (c)). Compared to the \noicm{} model, the additional energy inputs from the ICM enhance radiative cooling in all phases, including, while limited, cooling in the hot phase. We discuss the enhancement of X-ray luminosity as a function of RPS strengths (\autoref{sec:diss_phase}).

\subsection{Mixing Driven Acceleration and Stripping}\label{sec:stripping_mixing}

\begin{figure*}
    \centering
    \includegraphics[width=\textwidth]{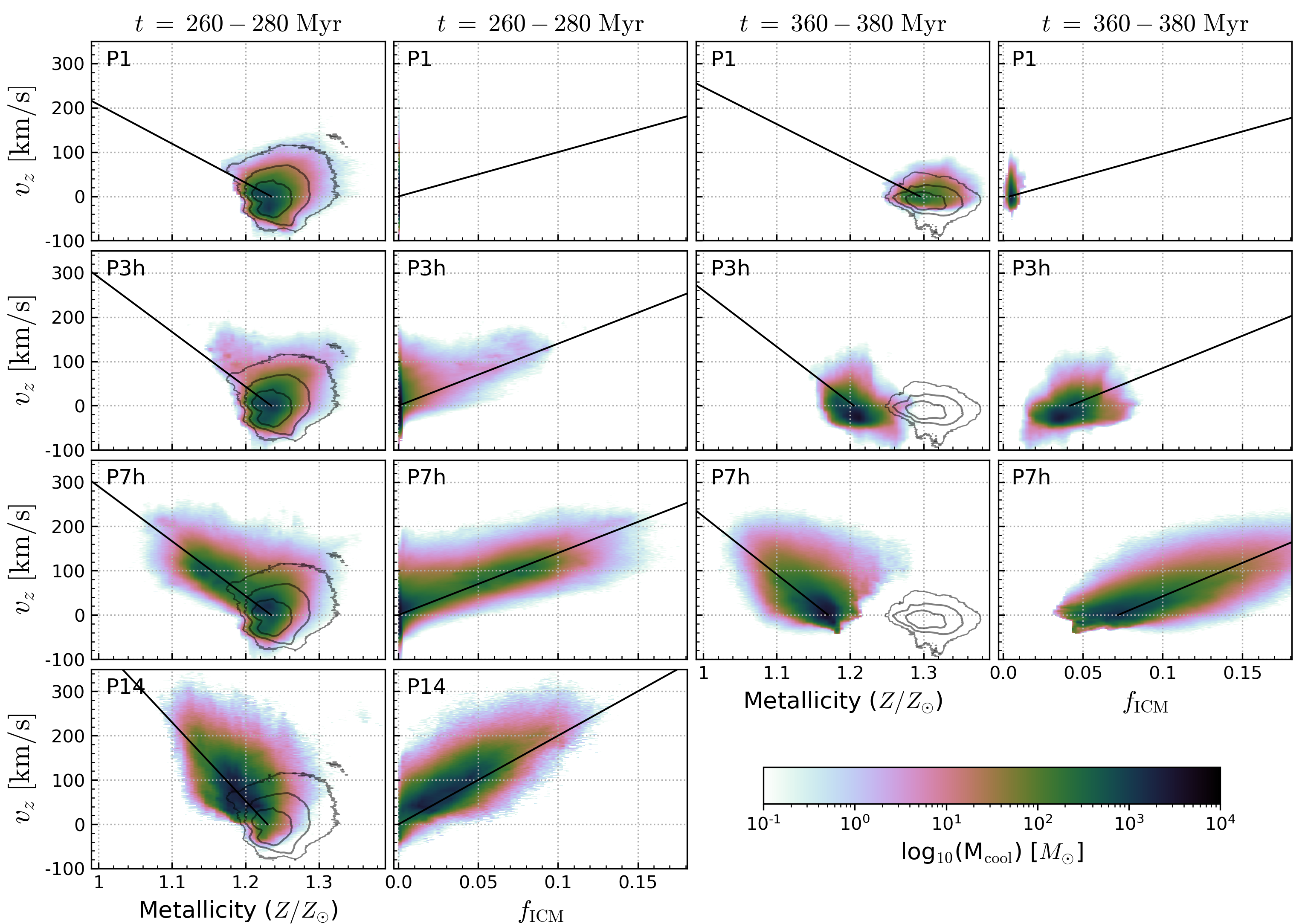}
    \caption{The mass distribution of the cool phase in the metallicity ($Z$) and vertical velocity ($v_z$) plane (1st and 3rd columns) and the ICM mass fraction $\sicm$ and vertical velocity ($v_z$) plane (2nd and 4th columns) over $z=1-2\kpc$. We consider two epochs at early times $t=260-280\Myr$ (left two columns) and late times $t=360-380\Myr$ (right two columns). From top to bottom, we show the models in ascending order of the ICM pressure, \icmpww{}, \icmpwh{}, \icmpsh{}, and \icmpss{}. As a reference, we show the mass distribution of the \noicm{} model in the the first and third columns as contours. In the second and fourth columns, we plot a linear relation between $v_z^{\rm cool}$ and $\sicm[cool]$ (\autoref{eq:vzcool_icm}). This can be translated into a linear relation between $v_z^{\rm cool}$ and $Z$ using \autoref{eq:Z} (assuming $\ssn=0$) as shown in the first and third columns.}
    \label{fig:svz_cool}
\end{figure*}

\begin{figure*}
    \centering
    \includegraphics[width=\textwidth]{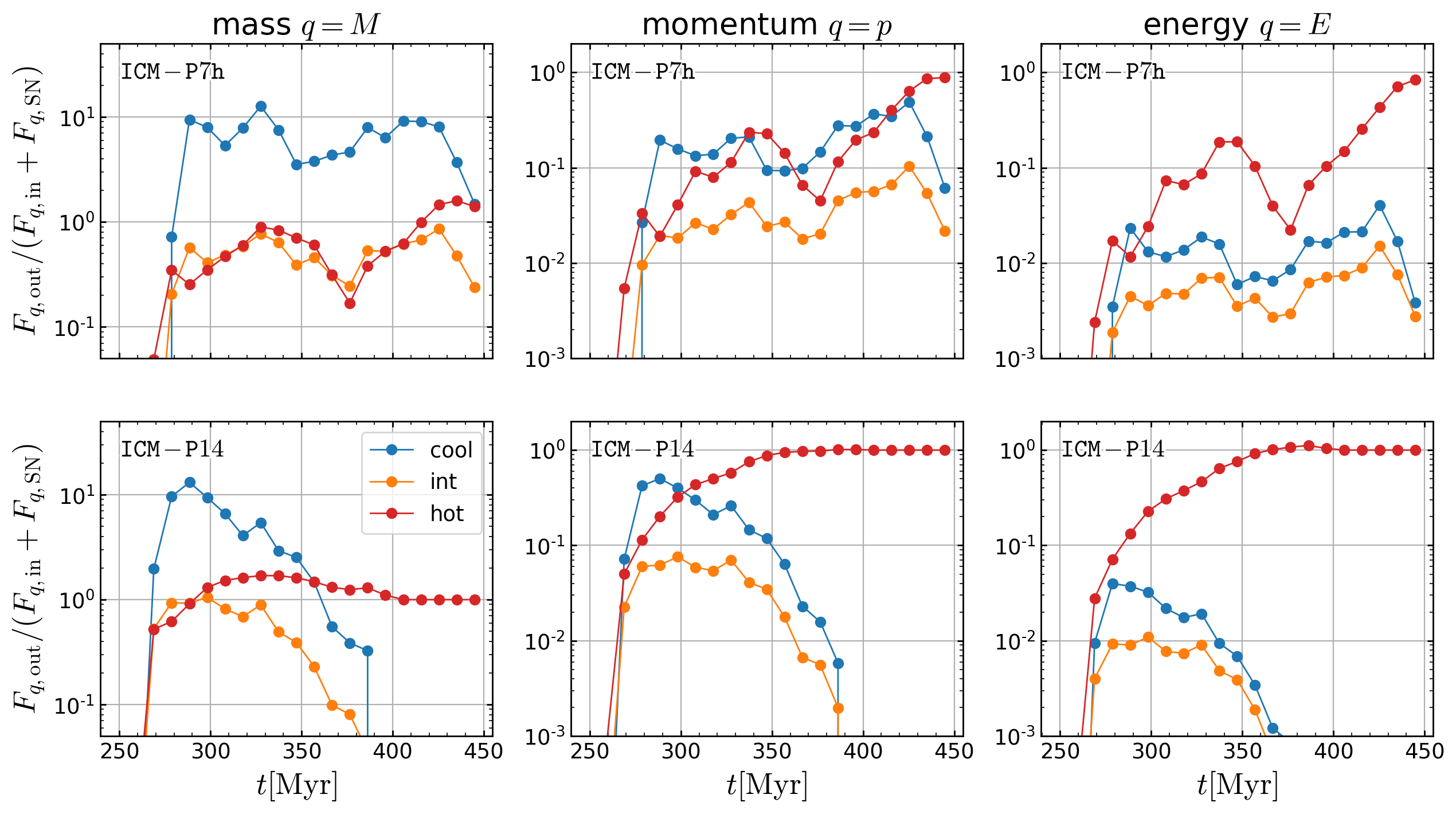}
    \caption{Outward fluxes of each phase through $z=3\kpc$ for the strong ICM models (top: \icmpsh{}; bottom: \icmpss{}) normalized by injected fluxes (by ICM and SNe). Colors of lines and symbols denote different thermal phases: blue for cool, orange for intermediate, and red for hot. The points denote the center of the time bin over which the time averaged outward flux (selecting only gas with $v_z>0$) is calculated using} \autoref{eq:Favg} with \autoref{eq:massflux} for mass flux, \autoref{eq:momflux} for momentum flux, and \autoref{eq:eneflux} for energy flux (neglecting the Poynting flux term). The injected ICM fluxes are similarly calculated at $z=-3\kpc$ for the gas with $v_z>0$. In the \icmpsh{} model (top), the outflow rates are more or less constant over time with significant mass carried out by the cool phase while the hot phase dominates energy flux. In contrast, the \icmpss{} model shows gradual decrease of the outgoing fluxes of the cool phase as the hot phase continuously shreds and entrains the cool phase in the hot outflows.
    \label{fig:out_in_strong}
\end{figure*}

Regarding the stripping of the ISM, the main question is how the cool ISM gets accelerated. If the cool ISM is accelerated as a whole by the drag force (or ram pressure force) from the ICM, the cool ISM simply gains outward momentum as is. In reality, the hot ICM penetrates through low-density channels in the multiphase ISM. Large relative velocities between the hot ICM and the cool ISM cause hydrodynamical instabilities, shredding the cool ISM and creating turbulent mixing layers. At the same time, strong radiative cooling due to the high cooling rate of the mixed gas in the mixing layers results in mass (and associated momentum and enthalpy) influx from the hot to cool gas. The competition between shredding and hot gas cooling leads to either net gain or loss of the cool gas mass. As the velocity associated with the mass gain (hot gas velocity) is in general significantly larger than the velocity associated with the mass loss (cool gas velocity), the cool ISM is accelerated by the momentum transfer associated with the mixing, while the direct acceleration still in play. 

As we show in \autoref{sec:phase_transition}, there is a significant phase transition occurs in the large regions of the ISM disk, especially in the strong ICM models, with corresponding momentum transfers seen in \autoref{sec:mom_transfer}. It is thus clear that the mixing must play a role. In this subsection, we seek evidence that the mixing-driven momentum transfer is actually a dominant mechanism for the cool ISM acceleration and stripping. 

The \emph{mixing-driven momentum transfer} simply means that the more the hot phase mixes in, the faster the cool phase moves. If this is the dominant mechanism, there must be a linear correlation between the velocity of the cool ISM and the mass fraction of the hot ICM (\autoref{eq:vzcool_icm}; see also \citealt{2020ApJ...895...43S,2021ApJ...911...68T}. The same is true for the SN-driven outflows. In our simulations, however, the SN ejecta tracer field is a less sensitive probe of the mixing-driven momentum transfer since the SN tracer field has been accumulated in all gas phases over many star formation-feedback cycles. On the other hand, the ICM mass fraction provides a telltale sign of the acceleration of the cool phase by the mixing of the hot ICM. Given the large difference in the metallicity of the ICM and ISM, this imprints a noticeable difference in the metallicity of the fast-moving cool ISM.

We choose two epochs ($t=260-280\Myr$ and $t=360-380\Myr$) and select the cool phase within $z=1-2\kpc$. \autoref{fig:svz_cool} plots the mass distribution in the metallicity $Z$ and vertical velocity $v_{z}^{\rm cool}$ plane (first and third columns) and the ICM mass fraction $\sicm[cool]$ and vertical velocity $v_z^{\rm cool}$ plane (second and fourth columns). The solid lines in the second and fourth columns are corresponding predictions from \autoref{eq:vzcool_icm}. The corresponding predictions for metallicity assuming $\ssn=0$ using \autoref{eq:Z} are shown in the first and third columns. For the fourth column, we apply an offset using the mean $\sicm[cool]$ as the baseline ICM fraction in the cool phase is nonzero and gradually increases. Except for the \icmpww{} model, all show tight correlations between the ICM mass fraction in the cool phase and the outward vertical velocity of the cool phase in the early epoch. The high-velocity component accelerated by the ICM appears as the low-metallicity component (anti-correlated component) in the $Z$-$v_z^{\rm cool}$ panels. Similarly, the mixing of the hot gas created by SNe produces a correlation between the metallicity excess and vertical velocity, but this correlation is less clear. The mixing-driven acceleration by the SN-origin hot gas is only visible at the early epoch of the \icmpwh{} model -- the high-velocity component correlated with metallicity $Z/Z_\odot\ge 1.2$. The contribution of the ICM-mixing increases as the ICM pressure increases, dominating the SN-origin mixing component. Generally, the outflow velocity is above the simple linear prediction, indicating the acceleration by ram pressure still contributes on top of the mixing-driven momentum transfer. The direct ram pressure drag is more important in the earlier evolution and the stronger ICM model when the ISM react to the ICM as a whole. At late times, cool clouds are more fragmented so that the crossection to the ICM inflows gets smaller while the surface area of the mixing layers gets larger.

In the late epoch (we omit the \icmpss{} model since there is almost no cool phase gas left at this epoch), a significant fraction of the cool phase gas falls back in the weak ICM models. Since this gas has been pushed upward mostly by the ICM previously, $\sicm[cool]$ is generally enhanced. The correlation between $\sicm[cool]$ and $v_z^{\rm cool}$ becomes less clear in the \icmpwh{} model due to the lack of continuous acceleration by the ICM but remains tight in the \icmpsh{} model.  At this epoch, the mean metallicity of the cool phase in both \icmpwh{} and \icmpsh{} models is greatly reduced compared to that of the \noicm{} model shown in contours. The metallicity of the cool phase decreases at least 0.1 dex in both models due to the ICM mixing.

We now ask, when the ISM is stripped by mixing with phase transition, which phase dominates in the outflows, hot (by shredding and escape before cooling) or cool (by cooling of stripped gas)? In \autoref{sec:phase_transition}, we show that the mass-dominating cool phase is shredded near the ICM-ISM interface and first stripped in the form of the intermediate and hot phases. Then, the significant cooling occurs within the simulation domain ($z\sim 2-3\kpc$) for the marginally strong ICM model (\icmps{}), but the majority of the stripped gas escapes the simulation domain before cooling in the strongest pressure model (\icmpss{}).

To provide a more quantitative view, we in \autoref{fig:out_in_strong} measure the outgoing fluxes ($F_{q,{\rm out}}$) normalized by the injected fluxes in the strong ICM models at $z=3\kpc$. The injected fluxes include the ICM inflow fluxes ($F_{q, {\rm in}}$) measured at $z=-3\kpc$ and those injected by SNe ($F_{q, {\rm SN}}$).\footnote{We calculate the mass, momentum, and energy fluxes from SNe using $\Delta N_{\rm SN}M_{\rm ej}/A\Delta t$, $\Delta N_{\rm SN} p_{\rm ref}/A\Delta t$, and $\Delta N_{\rm SN} E_{\rm SN}/A\Delta t$. For the momentum, we use the reference momentum $p_{\rm ref}=E_{\rm SN}/(2 v_{\rm cool})=1.25\times10^5\Msun \pc^{-2}$ of a SNR at the radiative stage \citep{2020ApJ...900...61K}, while SN ejecta mass $M_{\rm ej}=10\Msun$ and SN explosion energy $E_{\rm SN}=10^{51}\erg$ are our input parameters for SN feedback.}
Throughout the simulation, the contribution of SN feedback to total injected mass, momentum, and energy fluxes are 5, 28, and 17 \% for the \icmpsh{} model and 3, 14, and 6 \% for the \icmpss{} model, respectively. In the top row (\icmpsh{}), there is a significant and continuous outflowing mass, momentum, and energy fluxes from the cool phase. This is in stark contrast to the weak ICM models (and the \noicm{} model), where all outflow fluxes at $z=3\kpc$ are dominated by the hot phase at a level of a few \% of the injected fluxes \citep[e.g.,][]{2020ApJ...900...61K} and frequently truncated by inflows. The cool outflows in the \icmpsh{} model consist of both directly accelerated cool phase (mostly at early times) and additional cool phase created in the hot-to-cool phase transition layer (see \autoref{fig:mdot_tz}). Roughly 10-20\% of the injected momentum flux is transferred to the outflowing cool phase. The energy flux in the hot phase at this distance is about 10-20\% of the injected flux as significant thermal energy is transferred to the cooler phases and then radiated away. As a consequence, the energy flux in the cool and intermediate phases is much lower (a few percent) again because of large cooling losses (see \autoref{fig:Edot_tz_P7h}).

In the bottom row, strong outgoing fluxes in the cool phase exist only at very early times. The hot phase soon dominates all fluxes as the cool gas is shredded and evaporated to the hot phase. This \emph{mass loading} (or entrainment) to the hot outflowing gas increases the hot gas mass flux, at maximum, by a factor of two compared to the injected mass flux. Similar behavior is seen in the \icmpsh{} model at $t>400\Myr$. The maximum momentum transfer efficiency to the cool phase reaches up to 50\%. To summarize, both \icmpsh{} and \icmpss{} models show the mixing-driven ISM stripping soon after the direct cool phase acceleration, and the cool phase dominates the outflows in the early epoch. Then in \icmpss{}, the hot phase takes over the outflows whereas the cool phase outflow is maintained to some extent in \icmps{}.

\begin{figure}
    \centering
    \includegraphics[width=\columnwidth]{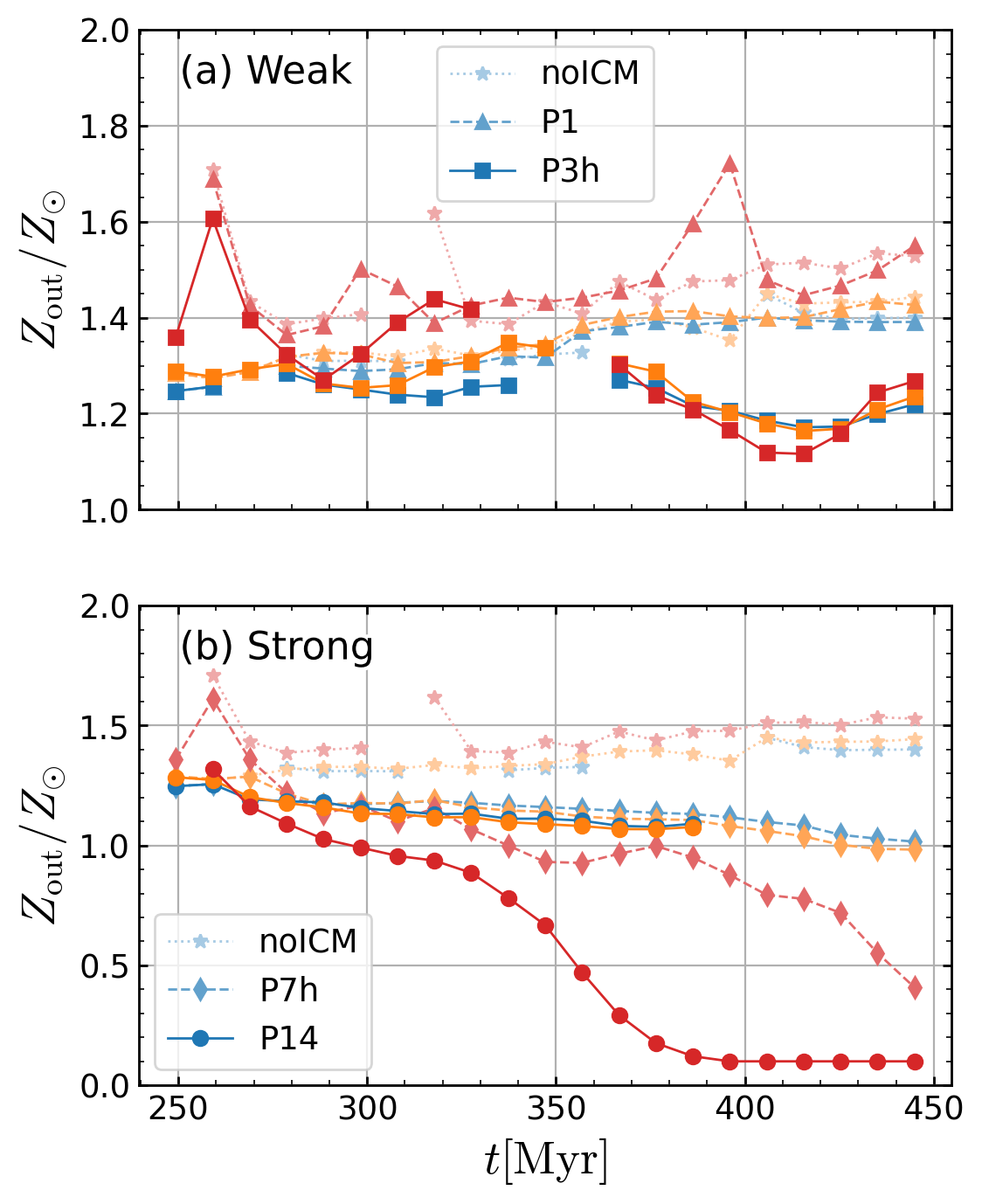}
    \caption{Outflow metallicity measured at $z=3\kpc$ for the (a) weak and (b) strong ICM models. Colors of lines and symbols denote different thermal phases: blue for cool, orange for intermediate, and red for hot. The same \noicm{} points are repeated in both panels.}
    \label{fig:Zout}
\end{figure}

Finally, we also make use of the metallicity of the outflowing gas to find the contribution of the ICM in accelerating the gas. \autoref{fig:Zout} plots the metallicity of the outflowing gas ($v_z>0$) at $z=3\kpc$ for the (a) weak and (b) strong ICM models along with the \noicm{} model in both panels as a reference. In the \icmpww{} model, the metallicities of the outflow in all three phases are essentially unchanged from those in the \noicm{} model. This is expected as the ICM cannot penetrate directly to the upper disk. The ICM is mixed into the ISM near the midplane, but the mass flux is insignificant. The outflowing gas is mostly driven out by SNe with enhanced metallicity.

The ICM makes a noticeable difference in the \icmpwh{} model. At $t\sim300-330\Myr$, the cool outflow metallicity is clearly reduced while the hotter phases still show metallicities similar to those in the \noicm{} and \icmpww{} models. The reduced metallicity in the cool outflow means the ICM mixing-driven acceleration as evidenced in \autoref{fig:svz_cool}. At later times ($t>350\Myr$), the outflow metallicity of all phases is significantly reduced, equally in all phases, and increases again. The decrease of metallicity indicates the mixing of the ICM into the ISM is the main driver of the outflows, while the later increase of the metallicity signals that the SN feedback plays a major role in driving outflows.

In the strong ICM models shown in \autoref{fig:Zout}(b), the metallicity of outflows is reduced at all times and keeps decreasing. This makes it plain that SNe in the strong ICM models is not a major driver of outflows, except very early time in the \icmpsh{} model as seen in \autoref{fig:svz_cool}. In \autoref{fig:Zout}(b), we also find that the outflow metallicity in the cool and intermediate phases is very similar for both strong ICM models, while the hot outflow metallicity is more reduced with stronger ICM pressure. The distribution of $\sicm[cool]$ shown in \autoref{fig:svz_cool} shows $\sicm[cool]<0.1-0.2$. Having demonstrated that the mixing is the main mechanism to drive outflows (or stripping), the limited range of the $\sicm[cool]$ implies that the outflowing cool gas would have been ablated and evaporated before mixing in too much hot ICM. The maximum ICM fraction then sets the maximum velocity and minimum metallicity difference of the cool gas accelerated by the ICM.

\section{Impact of the ICM on Star formation} \label{sec:sfr}

We take advantage of self-consistent modeling of star formation and feedback implemented in the TIGRESS framework to study the impact of the ICM ram pressure on star formation in and out of the ISM disks. We first present the changes in overall SFRs and their links to dense gas in the simulations in the presence of the ICM inflows. We then take a detailed look at extraplanar star formation.

\subsection{Enhancement and quenching of star formation}\label{sec:sfr_dichotomy}

\begin{figure*}[t]
\centering
\includegraphics[width=\textwidth]{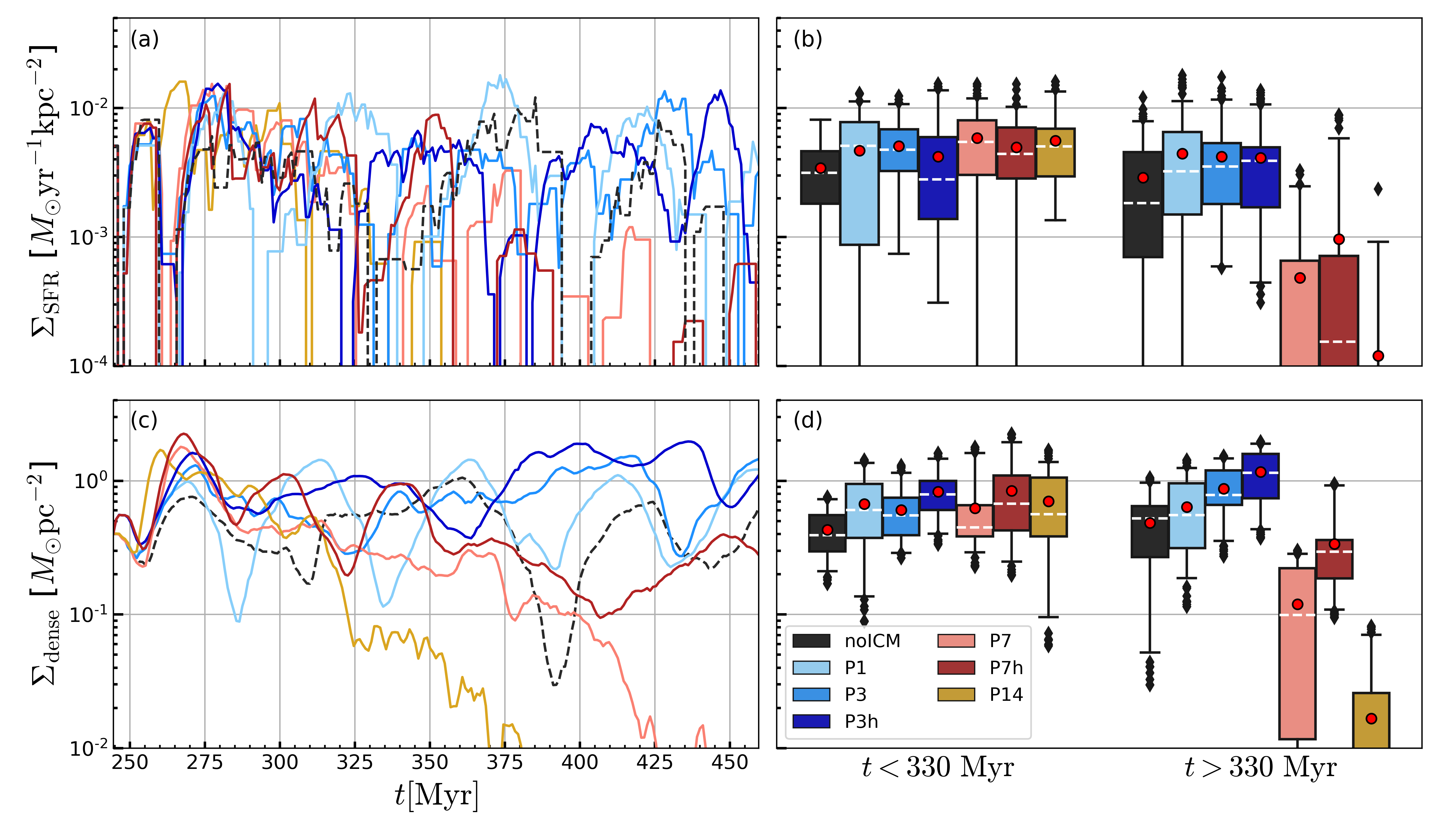}
\caption{\textbf{Left:} time evolution of (a) SFR surface density $\Sigma_{SFR}$ (\autoref{eq:SFR}) and (c) dense gas surface density $\Sigma_{\rm dense}\equiv \Sigma_{\rm gas}(n_H>10\pcc)$. The colored solid lines correspond to the models with the different ICM pressure, while the black dashed line is the \noicm{} model. \textbf{Right:} box and whisker plots of (b) $\Sigma_{SFR}$ and (d) $\Sigma_{\rm dense}$ for early ($t<330\Myr$) and late ($t>330\Myr$) periods. Boxes include 25th to 75th percentile with median (white dashed horizontal line) and mean (red circle). Whiskers represent 5th to 95th percentile with outliers shown as diamonds.}
\label{fig:sfr}
\end{figure*}

The general expectation is that strong ICM ram pressure that can strip the gas in galaxies will reduce SFRs in galaxies. At the same time, mild ICM ram pressure that compresses the gas in galaxies may enhance SFRs. Indeed, in our simulations, we find both effects depending on the ICM strengths and evolutionary stages. In short, SFRs are enhanced locally (inside the truncation radii) and temporarily (before active stripping), but the gas stripping eventually quenches star formation.

\autoref{fig:sfr}(a) plots the time evolution of SFR surface density defined by the total mass of young stars formed in the last $t_{\rm bin}=10\Myr$:
\begin{equation}\label{eq:SFR}
    \Sigma_{\rm SFR}(\Delta t=t_{\rm bin}) \equiv \frac{\Sigma m_{\rm sp}(t_{\rm m} < t_{\rm bin})}{L_xL_yt_{\rm bin}},
\end{equation}
where $m_{\rm sp}$ and $t_{\rm m}$ are the mass and mass-weighted mean age of the sink particle representing star clusters, respectively. This roughly corresponds to SFRs traced by H$\alpha$ \citep[e.g.,][]{2012ARA&A..50..531K}. \autoref{fig:sfr}(b) shows the box and whisker plots, presenting the distributions of $\Sigma_{\rm SFR}$ over two periods separated by $t=330$~Myr, before and after quenching of star formation in the strong ICM models. The enhancement of SFRs compared to the \noicm{} model in the early epoch is common in all models with the ICM. At later times ($t>330$~Myr), such enhancement of SFRs persists in the weak ICM models, while the gas stripping quenches star formation in the strong ICM models. The enhancement levels in $\Sigma_{\rm SFR}$ are $\sim 30\%$ to $50\%$ in the weak ICM models for more than 200 Myr explored in this paper. Also, the temporal modulation of $\Sigma_{\rm SFR}$ in these models gets stronger with higher peaks.

The enhancement of SFRs in the early epoch is mainly due to the compression of the overall ISM disk in the vertical direction. The introduction of the ICM inflows simply pushes the ISM from the lower disk to the midplane, effectively supplying more gas for star formation. In the weak ICM models, this \emph{additional} gas remains near the midplane where the majority of star formation takes place. However, strong ICM inflows can blow away the ISM altogether in $\sim 100\Myr$.  \autoref{fig:sfr}(c) and (d) show the time evolution and the box and whisker plots of the dense gas surface density $\Sigma_{\rm dense}$ selected by $n_H>10\pcc$. The first compression increases the peak dense gas mass at $\sim 270\Myr$ by about a factor of two in all ICM models. The corresponding enhancement of SFRs is delayed by $\sim 10\Myr$, a free-fall time of gas at $n_H=10\pcc$ \citep{2020ApJ...898...52M}. The enhancement of $\Sigma_{\rm dense}$ persists in the weak ICM models. In the strong ICM models, however, the dense gas mass quickly decreases over time due to shredding and stripping by the ICM. In the \icmps{} model, the dense gas still exists for a longer time than the \icmpss{} model. Some of the dense gas pushed far above the disk manages to form stars at late times (see \autoref{sec:extra_sf}).

\subsection{Extraplanar Star Formation}\label{sec:extra_sf}

\begin{figure*}[!ht]
\includegraphics[width=1\linewidth]{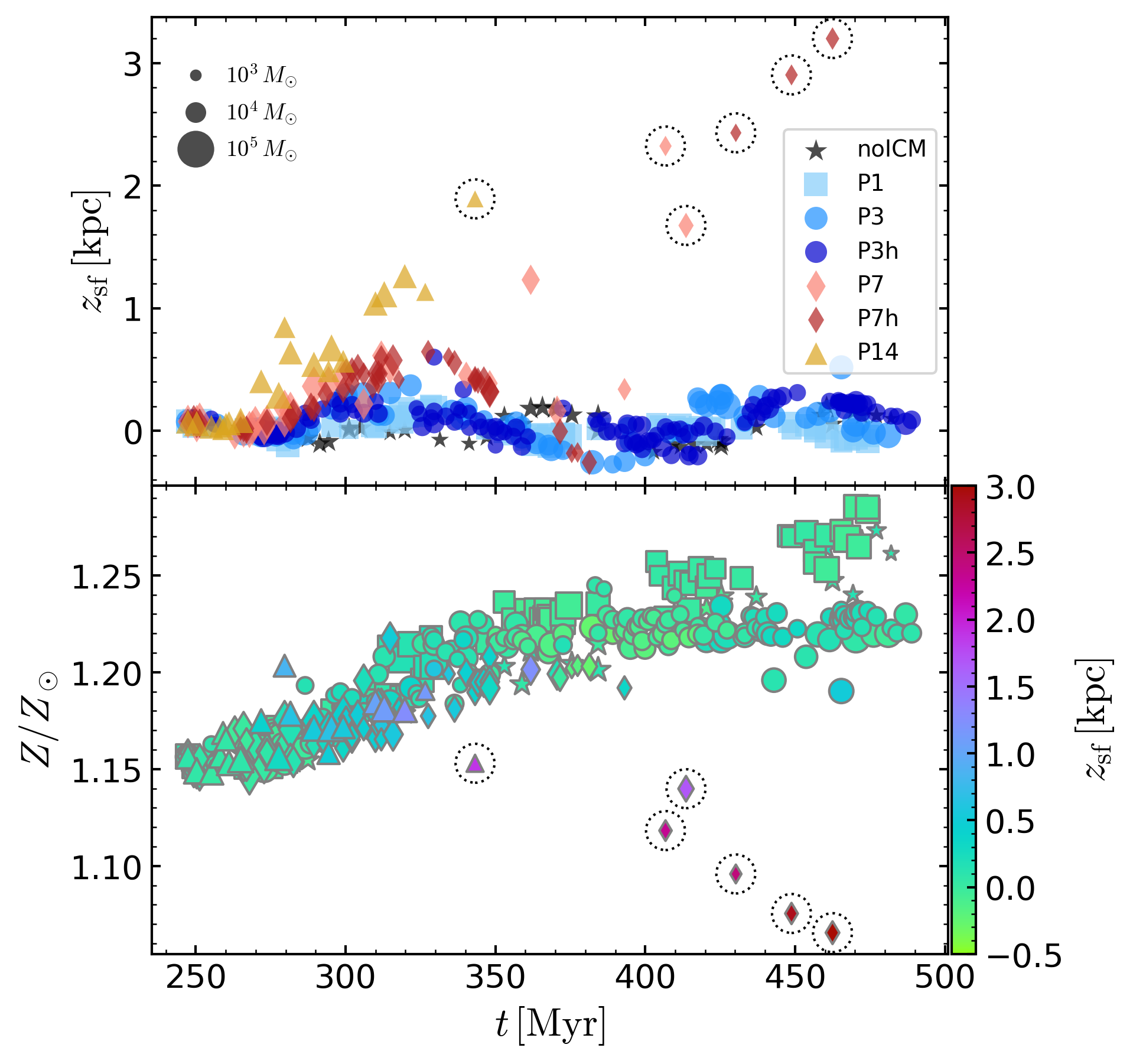}
\caption{{\bf Top:} vertical position at which new star clusters are born. The size of each symbol represents the mass of star cluster. {\bf Bottom:} metallicity of new star clusters colored by the star formation position. Star clusters with significant low metallicity are marked by block dotted circles in both panels.}
\label{fig:LOC}
\end{figure*}

One of the intriguing properties commonly found among the RPS galaxies is star-forming patches outside the stellar disk that remains intact. In \autoref{fig:LOC}(a), we show the vertical distance of the newly formed star clusters (sink particles) from the midplane $z_{\rm sf}$ over time. The size of symbols represents the mass of star clusters. The black star symbols are for the \noicm{} model.

On the one hand, for the strong ICM models, the bulk ISM keeps moving away from the midplane. As a consequence, $z_{\rm sf}$ increases over time. This continues for the \icmpss{} model, while the \icmps{} and \icmpsh{} models show turnover. Although one may get an impression that the ISM continuously moves upward in these models (see \autoref{fig:sicm}), the main gas reservoir is fragmented, and a large chunk of dense gas falls back (\autoref{fig:slc_late}(b)). As a result, two star-forming sites near and far from the midplane are visible at late times of the \icmps{} model. On the other hand, for the weak ICM models, as more and more gas moved the upper disk, the ISM weight shortly dominates the ICM pressure. The entire ISM disk falls back, and so does the star formation location. This introduces larger amplitude vertical oscillations of $z_{\rm sf}$ in the \icmpw{} and \icmpww{} models than the \noicm{} model in which a small amplitude vertical oscillation is naturally introduced by the asymmetry (\autoref{fig:noICM}).

\autoref{fig:LOC}(b) plots the metallicity of sink particles. Each symbol is now color-coded by $z_{\rm sf}$ shown in \autoref{fig:LOC}(a). In the \noicm{} model, the metallicity of new star clusters increases over time as the star-forming gas is continuously metal-enriched by mixing of the high-metallicity SN ejecta. The injected SN ejecta first goes into the hot phase and then quickly cools and mixes into the cool phase (see the top row of \autoref{fig:mdot_tz}). We find that the metallicity within the cool phase is nearly homogeneous, implying the efficient mixing of SN ejecta to the cold, star-forming gas. The metallicity of new stars born within the main ISM disk follows a common enrichment trend even with the ICM inflows, implying that they are born in the genuine ISM. However, in the strong ICM models, star clusters formed in the stripped gas far from the midplane at late times (marked by black circles) show lower metallicities compared to the enrichment trend. These star clusters are born in the gas that is experienced significant mixing with the low metallicity ICM. The gradual mixing of the ICM in the \icmpw{} and \icmpwh{} models also reduces the metallicity of new stars at late times ($t>400\Myr$), while higher SFRs with insignificant mass contribution from the ICM in the \icmpww{} model results in an even higher metallicity of new stars.

While reduced, the metallicity is still much higher than the ICM metallicity, implying that the composition of the star-forming gas in the extraplanar region is dominated by the genuine ISM. In our simulations, there is no sign of the complete shredding and recondensation of the star-forming cold gas in the stripped tails within the simulation domain $z<3.5\kpc$, which will be generally true for the extraplanar star formation within a few kpc away from the disks of RPS galaxies.

\section{Discussion}\label{sec:discussion}

\subsection{Ram Pressure Stripping as a Mixing-Driven Acceleration Process: Observational Imprints}

The multiphase nature of the ICM-ISM interaction is often neglected when developing theoretical understandings based on simple analytic models, although multiwavelength observations have revealed the multiphase gas involved in RPS galaxies such as cold molecular gas via CO \citep[e.g.,][]{2008A&A...491..455V,2015A&A...582A...6V,2018MNRAS.475.4055M,2017ApJ...839..114J,2019ApJ...883..145J}, cold and warm neutral gas via \ion{H}{1} \citep[e.g.,][]{1990AJ....100..604C,2009AJ....138.1741C,2010MNRAS.403.1175S},  warm ionized gas via H$\alpha$ \citep[e.g.,][]{fumagalli2014,boselli2016a}, and hot gas via X-rays \citep[e.g.,][]{sun2010,poggianti2019multiphase}. We show that the mass, momentum, and energy transfer from the hot ICM to the ISM via gas mixing is likely the dominant mechanism for stripping in our \emph{multiphase RPS} simulations. This is a qualitatively different process from that of a simple acceleration due to ram pressure without phase transition and mixing. A wealth of observational signatures will be imprinted on different gas phases.

\subsubsection{Imprints in Metallicity of Stripped Tails}

The main observational imprint of the mixing-driven acceleration model is on the anti-correlation of metallicity and cool gas velocity in the stripped tails (\autoref{fig:svz_cool}). If the ICM has distinctively lower metallicity than the ISM as we assumed, the fast-moving part of the stripped gas should have lower metallicity than the genuine ISM.\footnote{When SNe are the major source of the hot gas that mixes into the cool ISM, the metallicity in the fast-moving cool gas is likely enhanced compared to the genuine ISM \citep[see][]{2020ApJ...900...61K,2020ApJ...895...43S,2022ApJ...924...82F}.} Assuming $\ssn=0$, \autoref{eq:Z} and \autoref{eq:vzcool_icm} give the slope in the metallicity and velocity correlation $dZ/dv = (Z_{\rm ICM} - Z_{\rm ISM})/\vicm$. For $Z_{\rm ICM}/Z_{\rm ISM}=0.1$ and $\vicm=1000\kms$, this implies the metallicity reduction in the mixed cool gas is
\begin{equation}
    \frac{Z_{\rm mix}}{Z_{\rm ISM}}= 1 - 0.09 \rbrackets{\frac{\Delta v}{100\kms}},
\end{equation}
resulting in roughly 10\% reduction for 100~km/s difference in outflow velocity. The metallicity reduction decreases by a factor of two if $Z_{\rm ICM}/Z_{\rm ISM}=0.3$ and $\vicm=1400\kms$.

It is also noteworthy that the mixed ICM fraction in the cool gas cannot be arbitrarily high. Although efficient cooling helps to keep the mixed gas cool, the cool ISM can evaporate if the energy flux from the hot ICM is too large to be radiated away in the mixing layer. In our models, the difference of $\sicm$ between high and low-velocity cool gas is typically less than 0.1. This limits the dynamic range of outflow velocity less than $<200\kms$ even in \icmpss{} with $\vicm=2000\kms$. Similar results are also seen in numerical simulations of an isolated galaxy that is experiencing the hot ICM inflows \citep{2021ApJ...911...68T}, where they follow accelerated clouds in the stripped tails of 10s to 100s~kpc scales. Although the stripped tails traveled very far from the disk can have much larger $\sicm$ of clouds up to 0.8, the range of $\sicm$ at a particular position is limited to $\sim0.1-0.2$, translating into the velocity difference $\simlt 200-400\kms$. In addition to the maximum ICM fraction, there must be a significant fraction of total gas across a wide range of outflow velocities for such mixed gas to be visible. Since the gas mass fraction is sharply decreasing at high velocities in general \citep{2020ApJ...903L..34K}, observable velocity ranges and hence the metallicity differences can be further limited. Finally, it can possess a cleaner signal only if the stripping occurs quicker than the enrichment by SNe. Such condition favors strong ram pressure stripping galaxies. All the above makes the signal in gas phase metallicity difference at different outflow velocities we are searching for very small (of an order of 10\% or less). 

If the mixing-driven stripping continues beyond the immediately stripped tails we model here, there must be a well-defined trend in the mixed gas fraction as a function of distance across very long tails of RPS galaxies (often dubbed jellyfish galaxies). Indeed, global RPS galaxy simulations forming such long tails ($>100\kpc$) show a correlation between clouds' distance and ICM mass fraction \citep{2021ApJ...911...68T}. With an assumption that the mixing rate is constant over time, \citet{2021ApJ...911...68T} laid out a simple model for the ICM mass fraction as a function of distance, which qualitatively agrees with the increasing ICM mass fraction in clouds farther away in their simulations. The recent analysis of MUSE observations of RPS galaxies shows that warm ionized gas metallicities decrease as a function of distance from stellar disks \citep{2021ApJ...922L...6F}. The stripped gas, in reality, would experience much more complicated dynamical and thermal evolution, including deceleration by gravity, evaporation/fragmentation, and perhaps recondensation/growth by cooling in the mixing layers. 
The simple extrapolation of the clouds' velocity and ICM mass fraction at very far distances may not work well in predicting velocity and metallicity correlations quantitatively. Still, potentially illuminating results in \citet{2021ApJ...911...68T} (see their Figure 9) are that the slope in the cold clouds' velocity and ICM fraction correlation remains nearly linear over a large range of distances. Again, high precision measurements of metallicity across velocity channels to measure the slope in the $v$ and $Z$ correlation will be the most direct ways to confirm whether the mixing-driven acceleration is the dominant mechanism for the ram pressure stripping. 

RPS galaxies often show star formation activity outside the main, old stellar disk \citep[e.g.,][]{1999AJ....117..181K,sun2007,poggianti2016}. The mixing of the ICM also creates an imprint on the metallicity of stars formed in the extraplanar region. In the strong ICM models, the extraplanar star formation in the stripped tails occurs 2-3~kpc above the stellar disk, creating star clusters with lower metallicity (by  0.05--0.1 dexes) than those formed in the disk (see \autoref{fig:LOC}). Observationally, the relative difference of stellar metallicities between young stars from unstripped inner part and stripped extraplanar region can be compared. If the intrinsic metallicity gradient of galaxies is subtracted, stacking analysis may enhance a potential signal. 

\subsubsection{Imprints in Gas Phases}\label{sec:diss_phase}

RPS not only simply strips the cool ISM as is but also involves significant phase transition from cool to hotter phases (see \autoref{sec:stripping}). In the regions that experience strong RPS, the shredded cool gas escapes the simulation domain (stripped from galaxies) before it cools back (e.g., \icmpss{}). This is manifested in \ion{H}{1} deficiency in RPS galaxies \citep{2009AJ....138.1741C,2019MNRAS.487.4580R,2020A&A...640A..22R}. At the same time, the mass-loaded hot gas gets brighter in X-rays. The fate of such stripped gas is not traced in our simulations, but it is possible to cool back the stripped gas and form \ion{H}{1}/\Ha{} tails \citep[e.g.,][]{2019MNRAS.487.4580R,2020A&A...640A..22R}. In fact, large scale simulations do show late time cooling at more than tens of kpc away from the disk \citep{2012MNRAS.422.1609T,2021ApJ...911...68T,2022arXiv220101316L}, which might be responsible for the long, extended tails seen in \Ha{} and CO \citep[e.g.,][]{2008A&A...491..455V,2017MNRAS.466.1382L,2017ApJ...839..114J,2019ApJ...883..145J}.

Recently, sensitive, high-resolution observations of molecular gas tracers reveal the prevalence of extraplanar molecular gas in RPS galaxies \citep[e.g.,][]{2008A&A...491..455V,2018MNRAS.475.4055M,2017ApJ...839..114J,2019ApJ...883..145J}. The origin of extraplanar molecular clouds is unclear whether they are remnants of directly stripped molecular gas from the ISM disk or destroyed and reformed in the extraplanar region. \citet{2018ApJ...866L..10L} used high-resolution ALMA observations of NGC~4522 and detected ${}^{13}$CO in the extraplanar molecular clumps at $\lesssim$ a few kpc above the stellar disk near truncation radii. Given the relatively short formation time of molecular gas ($\sim10\Myr$) compared to the stripping time scale ($\sim100\Myr$), they concluded that both scenarios are feasible.

Our simulations lack the resolution and physical processes to follow the molecular species explicitly in the simulation. Instead, if we consider the dense gas ($n_H>10\pcc$; see \autoref{fig:sfr}) as a proxy of the molecular gas, we find that in the \icmpsh{} model where the later time extraplanar star formation occurs, (1) the dense gas fraction is comparable to the \noicm{} model and (2) most of the cold gas is located at 1-3~kpc above the midplane within which stars form at late times. This cold, dense gas is not completely shredded and recondensed, but a significant fraction of the ICM has been mixed in, which is evidenced by the lower metallicity of star clusters formed in the extraplanar regions (\autoref{fig:LOC}), up to $\sicm\sim 0.15$. Our simulations suggest that the extraplanar molecular gas (not those in the stripped tails farther than a few tens kpc from the disk) is mostly originated by the gas directly stripped from the disk. Still, the ICM mixing is important to accelerate the molecular gas (\autoref{fig:svz_cool}). Given that the \icmpss{} model quickly runs out its dense gas as the remaining enthalpy flux from the hot ICM after cooling is large enough to evaporate many small cold clouds, the marginally strong ICM condition can be optimal for pushing the molecular gas outward without destroying it. This translates into an expectation that the extraplanar molecular gas in active RPS galaxies can be most abundant near the truncation radii, which seems to be consistent with observations \citep{2018ApJ...866L..10L,2018MNRAS.475.4055M}. 

\begin{figure*}
    \centering
    \includegraphics[width=\textwidth]{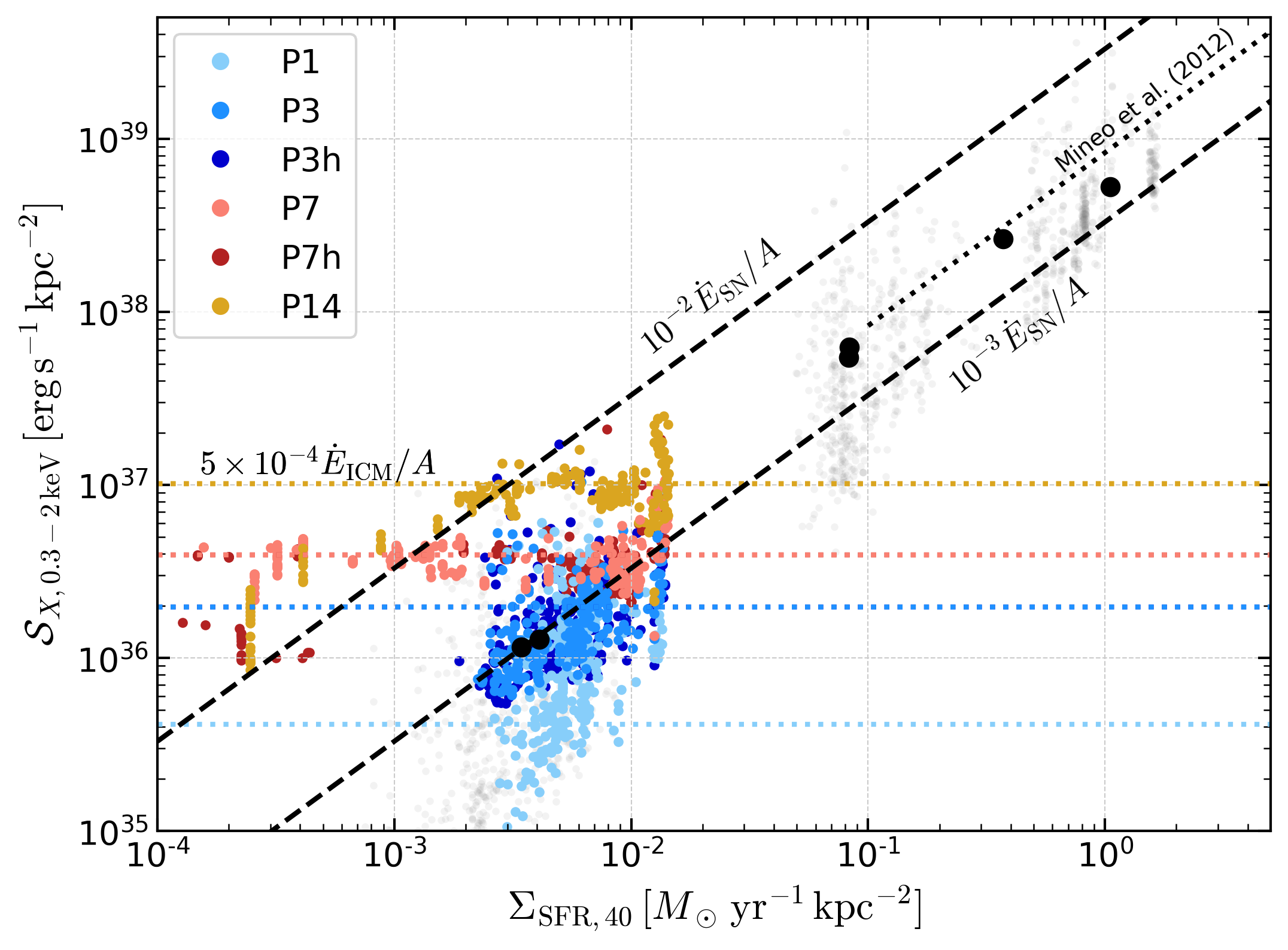}
    \caption{Soft X-ray surface brightness, $\mathcal{S}_{X,{\rm 0.3-2\,keV}}$, as a function of SFR surface density in the past 40~Myr, $\Sigma_{\rm SFR,40}$. The ICM models presented in this paper are shown as colored points, while the TIGRESS suite results (C.-G. Kim et al. in prep) are shown as gray points with their mean values as black circle for references. The two diagonal dashed lines denote the SN to soft X-ray efficiencies of $\epsilon_{\rm SN\rightarrow X}=1$ and 0.1\%. The horizontal dotted lines are for the ICM to soft X-ray conversion efficiency $\epsilon_{\rm ICM\rightarrow X}$ of 0.05\% for given ICM total energy flux.}
    \label{fig:Xray}
\end{figure*}

Another potentially strong observable signature of RPS can be an enhanced diffuse X-ray brightness \citep[e.g.,][]{2020MNRAS.494.5967K,2021ApJ...911..144C}. The diffuse thermal X-ray emission from the hot ISM is expected to correlate with SN rates and hence SFRs. From the observations of star forming galaxies, \citet{2012MNRAS.426.1870M} obtained a linear scaling relation between the diffuse (excluding resolved X-ray binaries) soft X-ray luminosity in $0.5-2$~keV ($L_{X,{\rm 0.5-2\,keV}}$) and SFR ($\dot{M}_*$) as $L_{X,{\rm 0.5-2\,keV}}/\dot{M}_* = 8.3\times10^{38}\ergs(M\odot\yr^{-1})^{-1}$ (see also \citealt{2003A&A...399...39R,2014MNRAS.437.1698M}). Assuming a canonical SN energy of $10^{51}\erg$, SN energy injection rate is $\dot{E}_{\rm SN} = 3.3\times10^{41}\ergs (\dot{M}_*/(M_\odot\yr^{-1}))$ for the standard initial mass function (1 SN per 100 $M_\odot$; e.g., \citealt{2001MNRAS.322..231K}). Then, the observed relation means the SN energy to soft X-ray conversion efficiency (or ``soft X-ray efficiency’’ in short) of $\epsilon_{\rm SN\rightarrow X}\equiv L_{X,{\rm 0.5-2\,keV}}/\dot{E}_{\rm SN}\sim0.25\%$.\footnote{\citet{2012MNRAS.426.1870M} calculated the intrinsic bolometric luminosity and derived the SN thermalization efficiency of $L_{\rm bol}/\dot{E}_{\rm SN}\sim 5\%$. But, the calculation of the intrinsic bolometric luminosity is largely model-dependent and requires many uncertain scaling factors. Here, we simply stick with the direct measurements of the diffuse soft X-ray luminosity and compare it with the forward-modelling results of our simulations.} The analysis of the TIGRESS suite \citep{2020ApJ...900...61K} shows similar soft X-ray efficiencies of $\epsilon_{\rm SN\rightarrow X}\sim0.1-0.2\%$ with higher efficiency for higher SFR surface density (C.-G. Kim et al. in prep; see \autoref{fig:Xray}).

Here, we calculate the X-ray surface brightness for the ICM models and compare them with the TIGRESS suite. We first obtain the X-ray emissivity in the soft X-ray band ($0.3-2$~keV) for each cell using the {\tt apec.v2} table (\url{http://www.atomdb.org};  \citealt{2012ApJ...756..128F}) adopted in {\tt yt} \citep{2011ApJS..192....9T}. The gas metallicity is fixed to solar metallicity. We then integrate total soft X-ray luminosity and divide it with the area to get the mean X-ray surface brightness. Figure~\ref{fig:Xray} plots as colored points the soft X-ray surface brightness $\mathcal{S}_{X,{\rm 0.3-2\,keV}}$ as a function of $\Sigma_{\rm SFR,40}$ (\autoref{eq:SFR} with $t_{\rm bin}=40\Myr$, the duration of SNe in each cluster). The weak ICM models are all consistent with the relation with soft X-ray efficiency of $0.1\%$ (lower dashed line) as the \noicm{} model in the TIGRESS suite. However, the strong ICM models show the enhancement of X-ray surface brightness consistent with a lower X-ray efficiency for the ICM inflows as $\mathcal{S}_{X,{\rm 0.3-2\,keV}}/(\dot{E}_{\rm ICM}/A) \sim 0.05\%$ (horizontal dotted lines), where the ICM total energy flux is $\dot{E}_{\rm ICM}/A = 0.5\dicm\vicm(\vicm^2+5\cicm^2)$. We note that the X-ray surface brightness of the pure ICM is much lower than that from the ICM-ISM interaction, mainly due to the ICM's low density. Somewhat lower X-ray efficiency of the ICM can be understood because of a larger mixing area involved in the ICM-ISM interaction than that of superbubbles driven by SNe. 

In RPS galaxies, the diffuse X-ray brightness can then be enhanced by an order of magnitude compared to that expected purely from SNe before the majority of the shock-heated gas is stripped away (shown as decreasing X-ray at lower SFRs in the strong RPS models).

\subsection{Ram pressure stripping and shock/wind-cloud interactions}\label{sec:rps_in_cloud_crushing}

The detailed look of the RPS process as multiphase gasdynamical interaction  is reminiscent of a collection of shock/wind-cloud interactions. The main question of shock/wind-cloud interaction studies is how cold gas can be accelerated before it is completely shredded. In adiabatic cases, the drag/acceleration time scale is always shorter than the cloud crushing time scale \citep{1994ApJ...420..213K}. In other words, the energy transferred from the hot shock/wind to the clouds is fully retained to heat up the clouds while surface instabilities shred them \citep[e.g.,][]{1994ApJ...420..213K,1994ApJ...433..757M,2009ApJ...703..330C,2015ApJ...805..158S,2015MNRAS.449....2M}. In order to prolong the cloud lifetime, several mechanisms have been proposed, including radiative cooling \citep[e.g.,][]{2009ApJ...703..330C} and magnetic fields \citep[in both wind and cloud, e.g,][]{2008ApJ...677..993D,2015MNRAS.449....2M,2020MNRAS.499.4261S}. Recently, it is realized that when clouds are large enough and cooling is strong, all the enthalpy flux can be radiated away while the significant mass and momentum of the hot phase are mixed and added into the cool phase without completely shredding the clouds. This allows the clouds to keep growing even while they are being accelerated by shock/wind-cloud interactions \citep[e.g.,][]{2016MNRAS.462.4157A,2018MNRAS.480L.111G,2020MNRAS.492.1970G,2021MNRAS.501.1143K,2020MNRAS.492.1841L,2019MNRAS.482.5401S,2020MNRAS.499.4261S}. It is certainly true that in our simulations, the chunk of the ISM that is facing the ICM inflows is large (a few 100 pc). The critical size above which the cool clouds can grow by cooling of the hot gas proposed by \citet{2018MNRAS.480L.111G} is
\begin{equation}
    R_{\rm crit}\approx 2\pc \frac{T_{\rm cl,4}^{5/2} \mathcal{M}_{\rm wind}}{P_3\Lambda_{\rm mix,-21.4}}\frac{\chi}{100},
\end{equation}
where $T_{\rm cl,4}\equiv T_{\rm cl}/10^4\Kel$ is the cloud temperature, $P_3 \equiv P/(10^3k_B\pcc\Kel)$ is the ambient pressure, $\mathcal{M}_{\rm wind}$ is the hot wind Mach number, $\Lambda_{\rm mix,-21.4}=\Lambda(T_{\rm mix})/(10^{-21.4}\ergs {\rm\,cm^{3}})$ is the cooling coefficient at the temperature of the mixed gas, and $\chi$ is the density contrast between wind and clouds. 

The critical size\footnote{The exact size criterion for the cool cloud growth is still under debate \citep{2021MNRAS.501.1143K}. \citet[][see also \citealt{2020MNRAS.499.4261S}]{2020MNRAS.492.1841L} suggest another criterion based on the hot gas cooling time and the predicted cloud lifetime from their simulations.} is of order of a few to tens of parsec at the typical conditions of our simulations with $P/k_B\sim 10^{4-5}\pcc\Kel$, $\chi=10^{2-3}$, and trans-to-subsonic wind Mach number (note that the ICM Mach number at injection was supersonic, but it is quickly thermalized and becomes subsonic at the time of interaction near the midplane). The bulk ISM from the first interaction cannot be completely shredded, while there are continuous shredding at the interfaces as the ICM penetrates through low density channels (see \autoref{fig:slc_early}). In the later time evolution, the strong ICM models successfully stripped the majority of the cool ISM from the disk midplane. There are smaller, fragmented cold cloudlets embedded in the ICM inflows (see \autoref{fig:slc_late}), which are vulnerable to shredding/evaporation. This results in a broad cold-to-hot phase transition layer seen in \autoref{fig:mdot_tz}(d) and (e). However, these clouds' wakes meet other cool gas and add their mass back to the cool phase. This evolution is more equivalent to that seen in shock-multicloud interactions \citep{2021MNRAS.506.5658B,2020MNRAS.499.2173B} rather than the growth of cool clouds in idealized shock/wind-cloud interaction simulations where the mass is added from an infinite hot reservoir via cooling of the mixed gas.

\subsection{Star formation in RPS galaxies}

The compression by the ICM can enhance SFRs by 30--50\% for a short period (a few tens of Myr), while the enhanced star formation is sustained in the weak ICM models (or inner part of an RPS galaxy) as the ISM remains compressed. In our strong models, representing the outer region of an RPS galaxy, star formation is quenched at time scales of $\sim100$ Myr.

Many previous simulations of RPS stripping galaxies including star formation \emph{recipes} commonly show the enhanced star formation activity before quenching \citep[e.g.,][]{2008A&A...481..337K,2012A&A...544A..54S,2017MNRAS.468.4107R}. Despite the qualitative agreements in the roles of RPS in star formation, global star formation enhancement found in many of these simulations is usually higher than ours by a factor of a few and persists longer \citep[][]{2008A&A...481..337K,2012A&A...544A..54S,2017MNRAS.468.4107R}. Keep in mind that earlier simulations in this category adopt a parameterized model for the ISM (cannot directly follow the gas phase cooler than $10^4\Kel$) and star formation \citep[e.g.,][]{1992ApJ...399L.113C,2003MNRAS.339..289S} to model an entire galaxy in a wind tunnel \citep[e.g.,][]{2008A&A...481..337K,2012A&A...544A..54S} or in a galaxy cluster \citep[e.g.,][]{2017MNRAS.468.4107R}. Therefore, the star formation rates obtained in previous global simulations can be sensitive to the adopted star formation recipes, although the global nature of such models (e.g., ICM wind inclination) can also be a reason for the difference (see \autoref{sec:caveats}).

Recently, \citet{2020ApJ...905...31L} presents simulations of an RPS galaxy with varying ICM inflow strengths and directions. Combined with adopted higher resolution (adaptive mesh refinement down to $20\pc$) and explicit ISM cooling and heating treatments, star formation in this work occurs in the cold, dense gas at number density above $100\pcc$, representing self-gravitating clouds, as in our simulations. In their moderate ICM inflow model, star formation at the outer region becomes suppressed during early 150 Myr, while the central region of the galaxy shows an enhancement of SFRs for several hundred Myr, compared to their {\tt NoWind} case. The quantitative agreements with our models are encouraging and indicative of the importance of high-resolution modeling of star formation in the multiphase ISM.

Star formation enhancement prior to the quenching has been observed in RPS galaxies \citep[e.g.,][]{2006ApJ...649L..75C,2014ApJ...780..119K}. Recently, \citet{2018ApJ...866L..25V} reported a systematic enhancement of the SFR (0.2 dex) for 42 RPS galaxies compared to the counterpart galaxies. The spatially resolved SFRs have been estimated for some of those galaxies, showing signs of central SFR enhancement before quenching \citep{vulcani2020_sfr_resolved}. In addition, \citet{roberts2020_coma_rps_sfr} identified 41 RPS candidate galaxies in the Coma cluster and reported enhanced SFR (0.3 dex) of them. Meanwhile, \citet[][]{2006ApJ...649L..75C, 2008AJ....136.1623C} measured the age of the youngest stellar population at the \ion{H}{1} truncation radii of RPS galaxies to estimate a quenching time scale -- how long ago star formation has been quenched since the \ion{H}{1} gas stripping. \citet{2008AJ....136.1623C} derived the quenching time scale of a few hundred Myr for the Virgo RPS galaxies which appear to be currently undergoing active RPS.
These results are broadly consistent with our results, while more spatially resolved analyses in observations are warranted for more quantitative comparisons.

\subsection{Caveats and Future Perspectives}\label{sec:caveats}

In this work, we had to limit our simulation domain to a kpc-size box to achieve high-resolution with explicit treatments of ISM physics \citep{2017ApJ...846..133K,2018ApJ...853..173K}. We thus cannot cover an entire galaxy nor model a galaxy orbiting within a realistic ICM. Consequently, we missed a few important physical processes involved in RPS.

First of all, we have to fix the ICM inflow direction perpendicular to the disk, i.e., face-on interaction. The ICM inflow inclination can be arbitrary for galaxies infalling/orbiting in a cluster. If the interaction is more edge-on, the ICM may preferentially compress the inflow-side ISM, while strips the extraplanar gas more easily  \citep{2006MNRAS.369..567R,2020ApJ...905...31L}. 

Second, the local model cannot capture global geometrical effects, which may be especially important in the stripping process at the truncation radius. After rapid stripping of gas outside the truncation radius, continuous stripping occurs through global hydrodynamical instabilities \citep[e.g.,][]{2005A&A...433..875R,2014ApJ...795..148T} in addition to local instabilities introduced by the penetrating ICM. Another interesting global effect is the inward radial migration of the stripped gas; the inner disk protects the tails from further interactions with the hot ICM. Such gas that is still bound to the galaxy will fall back \citep[e.g.,][]{2001MNRAS.328..185S,2009ApJ...694..789T,2014ApJ...795..148T}. In the future, more realistic, time-varying ICM inflows can be modeled, although global, cosmological models are needed for modeling of realistic variation of ICM ram pressure strengths and angles including the change of the orbits \citep[e.g.,][]{2019ApJ...874..161T}.

Third, although we vary the ICM ram pressure, we only consider a representative ISM disk condition similar to solar neighborhood. It is generally expected that the ratio of the ICM ram pressure to the ISM anchoring pressure $\Picm/\Wgg$ is the main control parameter that determines the dynamical impact of RPS. However, the microphysics of the ISM (e.g., chemistry and hence cooling and heating processes) will be particularly important for RPS in the multiphase ISM as the volume filling factors of different phase ISM vary over different conditions. The cooling rate in the mixing layer is one of the main parameter that determines the properties of mixing \citep{2021ApJ...911...68T,2022ApJ...924...82F}. In this regard, even for the same relative ram pressure strength, metallicity can change the efficiency of cooling and hence overall evolution of the multiphase RPS process.

The major advantage of the local framework used in this study is detailed modeling of ISM physics, which can be improved further by the future extensions of the TIGRESS framework with radiation and chemistry (J.-G. Kim et al. in prep). These capabilities are critically important to understand the evolution of the cold molecular gas \citep{2021MNRAS.505.1083G}. Although the current model follows gas at cold temperature $T<100\Kel$, within which star formation is modeled, the questions of the molecular cloud stripping remain unanswered. Are they intact during the journey to the far extraplanar region? How long do they survive? Can molecular clouds form again in the stripped tails? Future work using the new TIGRESS framework will shed light on these questions.

Finally, we point out the importance of thermal conduction, which is currently missing. As RPS should be viewed as the hydrodynamical interaction between gas phases at large temperature differences, conductive heat flux can be the dominant energy flux from the hot ICM to the ISM. The thermal conductivity of the ionized plasma increases steeply with the temperature \citep{1962pfig.book.....S}, but at the same time the conductive heat flux can be limited by the magnetic fields in the ISM that may wrap around the cool clouds as the ICM inflow sweeps up \citep[e.g.,][]{2008ApJ...678..274O}. Direct numerical simulations including anisotropic conduction within the TIGRESS framework where self-consistent magnetic field structure in the turbulent ISM is modeled are vital to understanding the role of thermal conduction in RPS galaxies.

\section{Conclusions}\label{sec:conclusion}

We conduct high-resolution MHD simulations of the ICM-ISM interaction to understand how the ICM strips the multiphase ISM from the galactic disk and how star formation changes in and out of the disk. We model the star-forming ISM using the TIGRESS numerical framework. We solve the ideal MHD equations in a local shearing-box with gas and (fixed) stellar gravity, optically thin cooling, star formation, and massive star feedback in the form of SNe and FUV radiative heating \citep{2017ApJ...846..133K}. We take a snapshot of a fully developed ISM from the solar-neighborhood model and simulate it with hot ICM inflows from the bottom boundaries to model face-on interactions of the disk ISM moving in a cluster. We adopt four different strengths of the ICM ram pressure $\Picm = \dicm \vicm^2$ while the ISM condition is fixed. The relative strength of the ICM ram pressure to the ISM anchoring pressure $\Wgg = 2\pi G \Sigma_{\rm gas} \Sigma_*$ covers a range of conditions representing the inner and outer radii of the truncation radius of a galaxy experiencing ram pressure stripping.

Our main findings are as follows:

\begin{enumerate}
    \item We find that the simple RPS condition comparing $\Picm$ and $\Wgg$ \citep{1972ApJ...176....1G} works well to predict overall stripping of the ISM disk even in our simulations with the multiphase ISM. Although the porous multiphase ISM structure allows the ICM to penetrate the disk through low-density channels and pollute the upper region of the disk regardless of the ICM strength, the effect of the ICM in accelerating the bulk ISM remains insignificant in the weak ICM models with $\Picm/\Wgg<1$. In this case, the majority of the ICM stays below the disk midplane (\autoref{fig:sicm}), and the gas remains within the ISM disk over the entire simulation duration ($\sim 250\Myr$). However, the ICM-ISM interface marches toward the other side of the ISM disk in the strong ICM models with $\Picm/\Wgg>1$. The ICM quickly strips the ISM in a half-mass stripping time scale of 60-130 Myr (\autoref{fig:surf}).
    
    \item In the strong ICM models, the mixing-driven momentum transfer from the ICM to the ISM plays an essential role in RPS (\autoref{sec:stripping}). At the ICM-ISM interface, the hot ICM inflow first shreds the cool ISM, adding mass into the hotter phases while all phases gain kinetic energy by ram pressure. In the stripped tails, the hot and intermediate phases (genuine ICM and shredded ISM) mix into the cool gas continuously. Most of the hot gas energy is radiated away, while mass and momentum are transferred to the cool phase. These hydrodynamical interactions between hot ICM (energy reservoir) and cool ISM (mass reservoir) result in accelerated cool gas after significant mixing of the hot ICM.
    
    \item The same momentum transfer process also occurs in the weak ICM models. But, the amount transferred to the ISM, together with the SN injected momentum fluxes, is simply used to support the deformed, one-sided disk with increased weight. There is not enough excess momentum and energy to drive strong, continuous outflows (or RPS) as in the strong ICM models.
    
    \item RPS via the mixing-driven momentum transfer imprints on the metallicity of the stripped tails. We find that star clusters formed in the stripped gas ($z>1\kpc$ from the midplane) show the metallicity lower than the new stars in the disk by $\sim 0.1$ dex (\autoref{fig:LOC}). Furthermore, we find a linear relationship between velocity and ICM mass fraction in the stripped cool gas as expected in the mixing-driven momentum transfer, giving rise to an equivalent anti-correlation between velocity and metallicity.
    
    \item Star formation is enhanced (30-50\%) in all ICM models at the early epoch of the simulation compared to the \noicm{} model. This enhancement persists in the weak ICM models for the entire simulation time ($\sim 250\Myr$), while the SFR is greatly reduced after $\sim$ 100 Myr in the strong ICM models.
\end{enumerate}

As the first results from novel RPS simulations using the local TIGRESS framework, we focus on general responses of the ISM disk with varying ICM inflow strengths. In a forthcoming paper, we will delve deep into the role of magnetic fields in the marginally strong model and the differential stripping of the cold and warm ISM. 

\acknowledgements

We acknowledge the anonymous referee for comments and suggestions that improved the clarity and quality of this paper.
AC and WC acknowledge support by the National Research Foundation of Korea (NRF), Grant No. 2018R1D1A1B07048314, 2022R1A2C100298211, and 2022R1A6A1A03053472.
C.-G.K. were supported by the National Aeronautics and Space Administration (NASA) through ATP Grant Number NNX17AG26G and Chandra Award Number TM0-21009X.
Resources supporting this work were provided in part by the NASA High-End Computing (HEC) Program through the NASA Advanced Supercomputing (NAS) Division at Ames Research Center and in part by the Princeton Institute for Computational Science and Engineering (PICSciE) and the Office of Information Technology’s High Performance Computing Center.

\newpage{}

\software{{\tt Athena} \citep{2008ApJS..178..137S,2009NewA...14..139S},
{\tt astropy} \citep{2013A&A...558A..33A,2018AJ....156..123T}, 
{\tt scipy} \citep{2020SciPy-NMeth},
{\tt numpy} \citep{vanderWalt2011}, 
{\tt IPython} \citep{Perez2007}, 
{\tt matplotlib} \citep{Hunter:2007},
{\tt xarray} \citep{hoyer2017xarray},
{\tt pandas} \citep{mckinney-proc-scipy-2010},
{\tt CMasher} \citep{CMasher},
{\tt adstex} (\url{https://github.com/yymao/adstex})
}


\bibliography{references,software}
\end{document}